\documentclass[prd,aps,twocolumn,a4paper,showkeys,nofootinbib,floatfix]{revtex4-1}

\usepackage{hyperref}
\usepackage{acronym}
\usepackage{booktabs}
\usepackage{dcolumn}

\usepackage{color,xcolor}
\usepackage{amsmath,amsfonts,amssymb}
\usepackage{mathrsfs}
\usepackage[utf8]{inputenc}
\usepackage[normalem]{ulem}
\usepackage{colortbl}
\usepackage{morefloats}

\usepackage{pifont} 

\usepackage{graphicx,psfrag}

\usepackage{multirow}
\usepackage{comment}

\definecolor{cyan}{rgb}{0,0.9,0.9}
\definecolor{orange}{rgb}{0.9,0.5,0}
\definecolor{magenta}{rgb}{1,0,1}
\definecolor{purple}{rgb}{0.8,0.4,0.8}
\definecolor{darkgreen}{rgb}{0.0,0.5,0.0}
\definecolor{gray}{rgb}{0.8242,0.8242,0.8242}
\definecolor{cadmiumgreen}{rgb}{0.0, 0.42, 0.24}

\def\arraybackslash{\let\\\tabularnewline}

\newcolumntype{d}[1]{D{.}{.}{#1}}

\newcommand{\ie}{\textit{i.e.}}
\newcommand{\eg}{\textit{e.g.,}}

\newcommand{\DSrefset}{\texttt{M0RefSet}} 

\newcommand{\DSheatcool}{\texttt{M0/M1Set}} 

\newcommand{\DScool}{\texttt{LeakSet}}

\newcommand{\DSnone}{\texttt{NoNusSet}}

\newcommand{\polql}{$P_2^2(q,\tilde{\Lambda})$}
\newcommand{\polq}{$P_2^1(\tilde{\Lambda})$}

\newcommand{\be}{\begin{equation}}
    \newcommand{\ee}{\end{equation}}
\newcommand{\bea}{\begin{eqnarray}}
    \newcommand{\eea}{\end{eqnarray}}
\newcommand{\bel}{\begin{align}}
    \newcommand{\eel}{\end{align}}

\newcommand{\cmark}{\ding{51}}
\newcommand{\xmark}{\ding{55}}

\DeclareOldFontCommand{\rm}{\normalfont\rmfamily}{\mathrm}
\DeclareOldFontCommand{\sf}{\normalfont\sffamily}{\mathsf}
\DeclareOldFontCommand{\tt}{\normalfont\ttfamily}{\mathtt}
\DeclareOldFontCommand{\bf}{\normalfont\bfseries}{\mathbf}
\DeclareOldFontCommand{\it}{\normalfont\itshape}{\mathit}
\DeclareOldFontCommand{\sl}{\normalfont\slshape}{\@nomath\sl}
\DeclareOldFontCommand{\sc}{\normalfont\scshape}{\@nomath\sc}
\DeclareRobustCommand*\cal{\@fontswitch\relax\mathcal}
\DeclareRobustCommand*\mit{\@fontswitch\relax\mathnormal}

\def\non{\nonumber}

\def\Msun{{\rm M_{\odot}}}
\def\GMc2{{\rm G M_{\odot} c^{-2}}}

\def\O{\mathcal{O}}

\def\md{M_{\rm ej}}
\def\vd{v_\infty}
\def\yd{Y_{e}}
\def\thetarms{\theta_{\rm RMS}}
\def\athetarms{\langle\thetarms\rangle}
\def\amd{\md}
\def\avd{\langle\vd\rangle}
\def\ayd{\langle\yd\rangle}

\def\chid{\chi^2 _{\nu}}
\def\ssr{\text{SSR}}

\begin{document}

\title{Mapping dynamical ejecta and disk masses
  from\\ numerical relativity simulations of neutron star mergers}
   
\author{Vsevolod \surname{Nedora}$^{1}$}
\author{Federico \surname{Schianchi}$^{2,1}$}
\author{Sebastiano \surname{Bernuzzi}$^{1}$}
\author{David \surname{Radice}$^{3,4,5}$} 
\author{Boris \surname{Daszuta}$^{1}$}
\author{Andrea \surname{Endrizzi}$^{1}$}
\author{Albino \surname{Perego}$^{6,7}$}
\author{Aviral \surname{Prakash}$^{3,4}$}
\author{Francesco \surname{Zappa}$^{1}$}

\address{${}^1$Theoretisch-Physikalisches Institut, Friedrich-SchillerUniversit\"{a}t Jena, 07743, Jena, Germany}
\address{${}^2$Institut f\"{u}r Physik und Astronomie, Universit\"{a}t Potsdam, Haus 28, Karl-Liebknecht-Str.  24/25,14476, Potsdam, Germany}
\address{${}^3$Institute for Gravitation \& the Cosmos, The Pennsylvania State University, University Park, PA 16802, USA}
\address{${}^4$Department of Physics, The Pennsylvania State University, University Park, PA 16802, USA}
\address{${}^5$Department of Astronomy \& Astrophysics, The Pennsylvania State University, University Park, PA 16802, USA} 
\address{${}^6$Dipartimento di Fisica, Universit\`{a} di Trento, Via Sommarive 14, 38123 Trento, Italy}
\address{${}^7$INFN-TIFPA, Trento Institute for Fundamental Physics and Applications, via Sommarive 14, I-38123 Trento, Italy}

\date{\today}

\begin{abstract}
  We present fitting formulae for the dynamical ejecta properties and remnant disk
  masses from the largest to date 
  sample of numerical relativity simulations.
  The considered data include some of the latest simulations with
  microphysical nuclear equations of state (EOS) and neutrino transport as well as other results with
  polytropic EOS available in the literature. 
  Our analysis indicates that the broad features of the dynamical ejecta and disk properties can be  
  captured by fitting expressions, that depend on mass ratio and reduced tidal parameter. 
  The comparative analysis of literature data shows that 
  microphysics and neutrino absorption have a significant impact on the
  dynamical ejecta properties. Microphysical nuclear EOS lead to  
  average velocities smaller than polytropic EOS, while
  including neutrino absorption results in larger average ejecta masses
  and electron fractions.
  Hence, microphysics and neutrino transport are necessary 
  to obtain quantitative models of the ejecta in terms of the binary
  parameters.
\end{abstract}


\maketitle

\section{Introduction}

The UV/optical/NIR transient AT2017gfo
\cite{Arcavi:2017xiz,Coulter:2017wya,Drout:2017ijr,Evans:2017mmy,Hallinan:2017woc,Kasliwal:2017ngb,Nicholl:2017ahq,Smartt:2017fuw,Soares-santos:2017lru,Tanvir:2017pws,Troja:2017nqp,Mooley:2018dlz,Ruan:2017bha,Lyman:2018qjg}, 
counterpart of the gravitational-wave signal GW170817
\cite{Abbott:2017wuw,Abbott:2017oio,Abbott:2018wiz,Abbott:2018hgk}, is
explained as the kilonova signal from the radioactive decay of
$r$-process elements synthesized in the mass ejected during binary neutron
star mergers \cite{Lattimer:1974a,Symbalisty:1982a,Li:1998bw,Kulkarni:2005jw,Rosswog:1998hy,Rosswog:2005su,Metzger:2010sy,Roberts:2011xz,Kasen:2013xka,Cowperthwaite:2017dyu,Villar:2017wcc,Tanvir:2017pws,Tanaka:2017qxj,Perego:2017wtu,Kawaguchi:2018ptg}. 
Minimal models of the kilonova AT2017gfo
require at least two ejecta components to account for the observed light curves: a lanthanide-poor (for the
blue signal) and a lanthanide-rich (for the red signal) one
\cite{Cowperthwaite:2017dyu,Villar:2017wcc,Tanvir:2017pws,Tanaka:2017qxj,Perego:2017wtu,Kawaguchi:2018ptg}. 
These components are often identified as the dynamical ejecta and the wind ejecta
from the remnant disk, although simulations clearly indicate that this
interpretation is not complete. \eg~\cite{Fahlman:2018llv,Nedora:2020pak}

Mass ejection in mergers can be triggered by different
mechanisms acting on different timescales (see
\cite{Metzger:2019zeh,Shibata:2019wef,Radice:2020ddv,Bernuzzi:2020tgt} for reviews on various aspects of the problem). 
Simulations robustly identify dynamical ejecta, of mass $\md\sim\O(10^{-4}-10^{-2})\,\Msun$ launched during
merger at average velocities $\avd\sim0.1-0.3\,$c, \eg~
\cite{Rosswog:1998hy,Rosswog:2005su,Hotokezaka:2013iia,Bauswein:2013yna,Wanajo:2014wha,Sekiguchi:2015dma,Radice:2016dwd,Sekiguchi:2016bjd,Vincent:2019kor},
and (for many fiducial postmerger configurations) more massive but slower winds launched on secular timescales from the remnant disk~\cite{Dessart:2008zd,Fernandez:2014bra,Perego:2014fma,Just:2014fka,Kasen:2014toa,Metzger:2014ila,Martin:2015hxa,Wu:2016pnw,Siegel:2017nub,Fujibayashi:2017puw,Fahlman:2018llv,Metzger:2018uni,Fernandez:2018kax,Miller:2019dpt}.
The most accurate approach to compute the dynamical ejecta and the
remnant evolution is to employ ab-initio 3+1 simulations in numerical relativity,
\eg~\cite{Hotokezaka:2012ze,Hotokezaka:2013iia,Wanajo:2014wha,Sekiguchi:2015dma,Dietrich:2015iva,Palenzuela:2015dqa,Bernuzzi:2015opx,Radice:2016dwd,Lehner:2016lxy,Sekiguchi:2016bjd,Radice:2018pdn,Vincent:2019kor,Perego:2019adq,Kiuchi:2019lls,Endrizzi:2019trv,Bernuzzi:2020txg}.
The increasing amount of data (both in terms of simulated binaries,
physics input and numerical resolutions) allows us to explore the dependencies of ejecta and
remnant properties on the binary parameters.
Fitting formulae of numerical relativity data for the dynamical ejecta and remnant disk properties 
from binary neutron star mergers have been previously presented in
\cite{Dietrich:2016fpt,Radice:2018pdn,Kruger:2020gig}. 
The interest in these formulae is at least twofold. On the one hand, they can be used to identify the main
parametric dependencies of the ejecta mechanisms; on the other hand, they can be employed to constrain the source parameters from kilonova observations,
\eg~\cite{Radice:2017lry,Perego:2017wtu,Coughlin:2018fis,Coughlin:2019zqi}. 
Additionally, they are key to predict the amount and the properties of the ejecta that enter chemical evolution models, \eg \cite{Bonetti:2019fxj}.

Here we employ an extended set of data presented in previous works
that includes also recent simulations with approximate neutrino
transport and large mass ratios
\cite{Perego:2019adq,Nedora:2019jhl,Bernuzzi:2020txg,Nedora:2020pak}. 

We re-calibrate the fit models proposed in the literature with this extended dataset. 
Additionally we test simple polynomials as fitting models for the ejecta mass,
velocity, and electron fraction.

\begin{table*}[t]
    \caption{
        Datasets with the dynamical ejecta data and disk masses
        employed in this work. The available data is shown in the columns
        starting from the fourth, that contain: gravitational mass of the binary, baryonic mass of
        the binary, reduced tidal parameter, ejecta mass, ejecta
        velocity, ejecta electron fraction, disk/torus mass. EOS are
        either microphysical or piecewise polytropic (PWP). Neutrino
        schemes are: leakage, leakage + M0 or M1 for free streaming
        neutrinos, or M1. 
        The compiled data are available online at \citep{vsevolod_nedora_2020_4283517}.
    }
    \label{tab:data}
    \begin{tabular}{ccccccccccc}
        \hline\hline
        Ref.  & EOS  & Neutrinos & $M$  & $M_b$  & $\tilde{\Lambda}$ & $M_{\rm ej}$ & $\upsilon_{\rm ej}$ & $Y_e$  & $M_{\rm disk}$ & Dataset
        \\ \hline \hline
        \multicolumn{1}{c|}{\cite{Perego:2019adq}}     & \multicolumn{1}{c|}{Micro} & Leak+M0    & \cmark & \cmark & \cmark & \cmark & \cmark  & \cmark & \cmark &  \DSrefset{} \& \DSheatcool \\
        \multicolumn{1}{c|}{\cite{Nedora:2019jhl}}     & \multicolumn{1}{c|}{Micro} & Leak+M0    & \cmark & \cmark & \cmark  & \cmark  & \cmark & \cmark & \cmark  & \DSrefset{} \& \DSheatcool  \\
        \multicolumn{1}{c|}{\cite{Bernuzzi:2020txg}}   & \multicolumn{1}{c|}{Micro} & Leak+M0    & \cmark & \cmark & \cmark & \cmark & \cmark & \cmark & \cmark & \DSrefset{} \& \DSheatcool   \\
        \multicolumn{1}{c|}{\cite{Nedora:2020pak}}        & \multicolumn{1}{c|}{Micro} &   Leak+M0   & \cmark & \cmark & \cmark & \cmark & \cmark & \cmark & \cmark   & \DSrefset{} \& \DSheatcool   \\
        \hline
        \multicolumn{1}{c|}{\cite{Vincent:2019kor}}    & \multicolumn{1}{c|}{Micro}  &  M1  & \cmark & \cmark & \cmark & \cmark & \cmark  & \cmark & \xmark   &  \DSheatcool{} \\
        \multicolumn{1}{c|}{\cite{Sekiguchi:2015dma}}  & \multicolumn{1}{c|}{Micro} &  Leak+M1   & \cmark & \xmark & \xmark  & \cmark & \xmark & \cmark & \xmark &  \DSheatcool \\
        \multicolumn{1}{c|}{\cite{Sekiguchi:2016bjd}}  & \multicolumn{1}{c|}{Micro} &  Leak+M1   & \cmark & \xmark & \xmark & \cmark & \cmark & \cmark & \cmark & \DSheatcool \\
        \multicolumn{1}{c|}{\cite{Radice:2018pdn}~(M0)} & \multicolumn{1}{c|}{Micro} & Leak+M0    & \cmark & \cmark & \cmark & \cmark & \cmark & \cmark & \cmark  &  \DSheatcool   \\
        \hline
        \multicolumn{1}{c|}{\cite{Lehner:2016lxy}}     & \multicolumn{1}{c|}{Micro} & Leak    & \cmark & \cmark & \xmark & \cmark & \cmark & \xmark & \xmark   &  \DScool \\
        \multicolumn{1}{c|}{\cite{Radice:2018pdn}~(LK)} & \multicolumn{1}{c|}{Micro} & Leak    & \cmark & \cmark & \cmark & \cmark & \cmark & \cmark & \cmark  & \DScool   \\
        \hline
        \multicolumn{1}{c|}{\cite{Kiuchi:2019lls}}     & \multicolumn{1}{c|}{PWP}   &  -    & \cmark & \cmark & \cmark & \cmark & \xmark  & \xmark & \cmark    & \DSnone \\
        \multicolumn{1}{c|}{\cite{Dietrich:2016hky}}   & \multicolumn{1}{c|}{PWP}  &  -     & \cmark & \cmark & \cmark & \cmark & \cmark  & \xmark & \cmark  &  \DSnone \\
        \multicolumn{1}{c|}{\cite{Dietrich:2016hky}}   & \multicolumn{1}{c|}{PWP}  &   -    & \cmark & \cmark & \cmark & \cmark & \cmark & \xmark & \cmark  & \DSnone \\
        \multicolumn{1}{c|}{\cite{Hotokezaka:2012ze}}  & \multicolumn{1}{c|}{PWP}  &    -   & \cmark & \xmark & \xmark & \cmark & \cmark & \xmark & \xmark  &  \DSnone \\
        \multicolumn{1}{c|}{\cite{Bauswein:2013yna}}   & \multicolumn{1}{c|}{Micro}&  - & \cmark & \xmark & \xmark & \cmark & \cmark & \xmark & \xmark  &  \DSnone \\
        \hline\hline
    \end{tabular}
\end{table*}

Throughout the paper we label the two NSs with subscripts $A$, $B$. 
The individual gravitational masses are indicated as $M_A$, $M_B$, 
the baryonic masses as $M_{b~A}$, $M_{b~B}$, 
the total mass as $M = M_A + M_B$, 
and the mass ratio $q=M_A/M_B\geq1$. 

We define the quadrupolar tidal parameters as
$\Lambda_i \equiv 2/3\, C_i^{-5} k^{(2)}_i$
where $k_i^{(2)}$ is the dimensionless gravitoelectric Love number \cite{Damour:2009vw}, 
$C_i \equiv GM_A/(c^2R_A)$ the compactness parameter, and $i=A,B$.
The reduced tidal parameter \cite{Favata:2013rwa} is:
\be
\tilde\Lambda = \frac{16}{13}\frac{(M_A+12 M_B)M_A^4 \Lambda_A}{M^5}+(A\leftrightarrow B)\,.
\label{eq:Lambda_tilde}
\ee
We use CGS units except for masses and velocities, given in units of $\Msun$ and $c$, respectively.

\section{Data and Method}\label{sec:method}

The datasets used in this paper are summarized in Tab.~\ref{tab:data}.
We group them with respect to the employed neutrino treatment:

\begin{itemize}
    
    \item \DSheatcool{} comprises a set of models with neutrino emission 
    and absorption and microphysical EOS. It includes 
    $8$ models with leakage+M0 of \cite{Radice:2018pdn} and models 
    of \cite{Sekiguchi:2015dma,Sekiguchi:2016bjd,Vincent:2019kor}
    in which a leakage+M1 scheme or a M1 gray scheme are employed for the neutrino transport. 
    Models reported in these works span 
    $q\in[1, 1.30]$, 
    $\tilde{\Lambda}\in[340, 1437]$, 
    $M_{\rm tot}\in[2.52,2.88]$, 
    and $M_{\rm chirp}\in[1.10,1.25]$.
    
    \item \DSrefset{} harbors models with similar physical setup as those 
    of \DSheatcool{} (specifically, they were computed with the same setup as 
    models with leakage+M0 neutrino scheme of \cite{Radice:2018pdn}).
    Presented in \cite{Perego:2019adq,Nedora:2019jhl,Bernuzzi:2020txg,Nedora:2020pak} 
    these models are uniform in terms of the 
    numerical setup, code and physics and have fixed chirp mass. 
    For that reason we group them into a separate, reference dataset. 
    The models of this set span $q\in[1, 1.82]$, 
    $\tilde{\Lambda}\in[400, 850]$, 
    $M_{\rm tot}\in[2.73,2.88]$ with 
    the chirp mass $M_{\rm chirp}=1.19$.
    
    \item \DScool{} comprises models with leakage scheme as neutrino treatment and microphysical EOS.
    The dataset includes a subset of models from \cite{Radice:2018pdn} ($35$ runs denoted as LK),
    and the set of models from \cite{Lehner:2016lxy}.
    The models in this dataset span $q\in[1, 1.31]$, 
    $\tilde{\Lambda}\in[116, 1688]$, 
    $M_{\rm tot}\in[2.40,3.42]$, 
    and $M_{\rm chirp}\in[1.04,1.49]$.
    
    \item \DSnone{} is composed of models with piecewise-polytropic EOSs 
    \cite{Hotokezaka:2012ze,Dietrich:2015iva,Dietrich:2016hky,Kiuchi:2019lls,Bauswein:2013yna},
    in which temperature effects are approximated by a
    gamma-law pressure contribution, while
    composition and weak effects are neglected.
    The models in this dataset span 
    $q\in[1, 2.06]$, 
    $\tilde{\Lambda}\in[50, 3196]$, 
    $M_{\rm tot}\in[2.4,4.0]$, 
    and $M_{\rm chirp}\in[1.04,1.74]$.
    
\end{itemize}

\begin{figure*}[t]
    \centering 
    \includegraphics[width=0.32\textwidth]{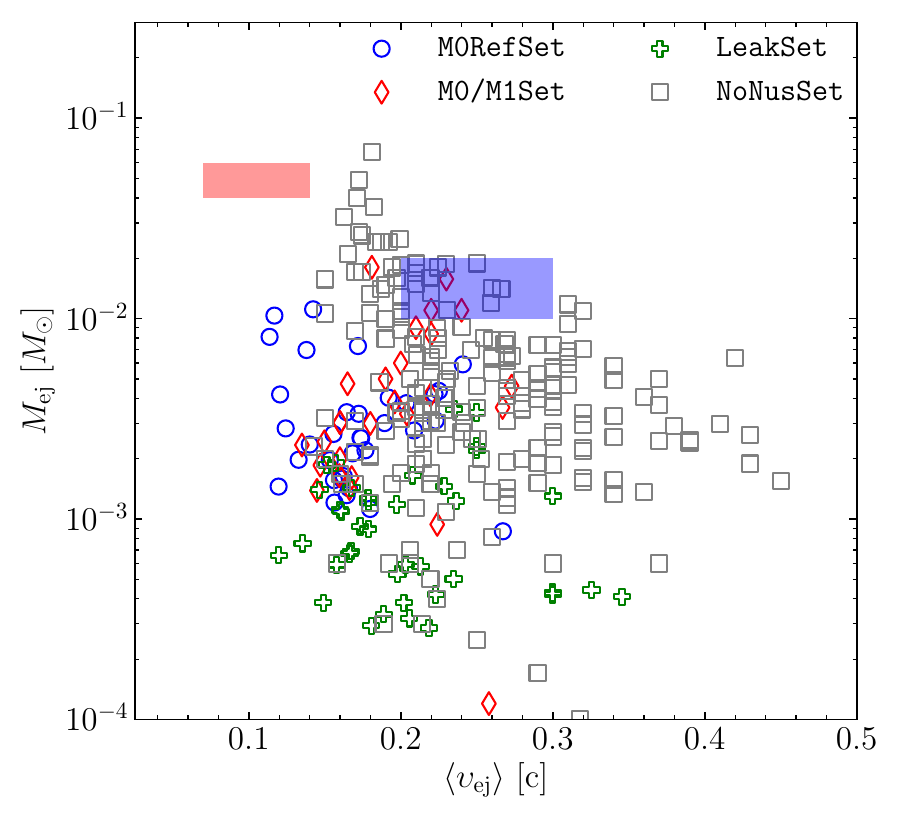}
    \includegraphics[width=0.32\textwidth]{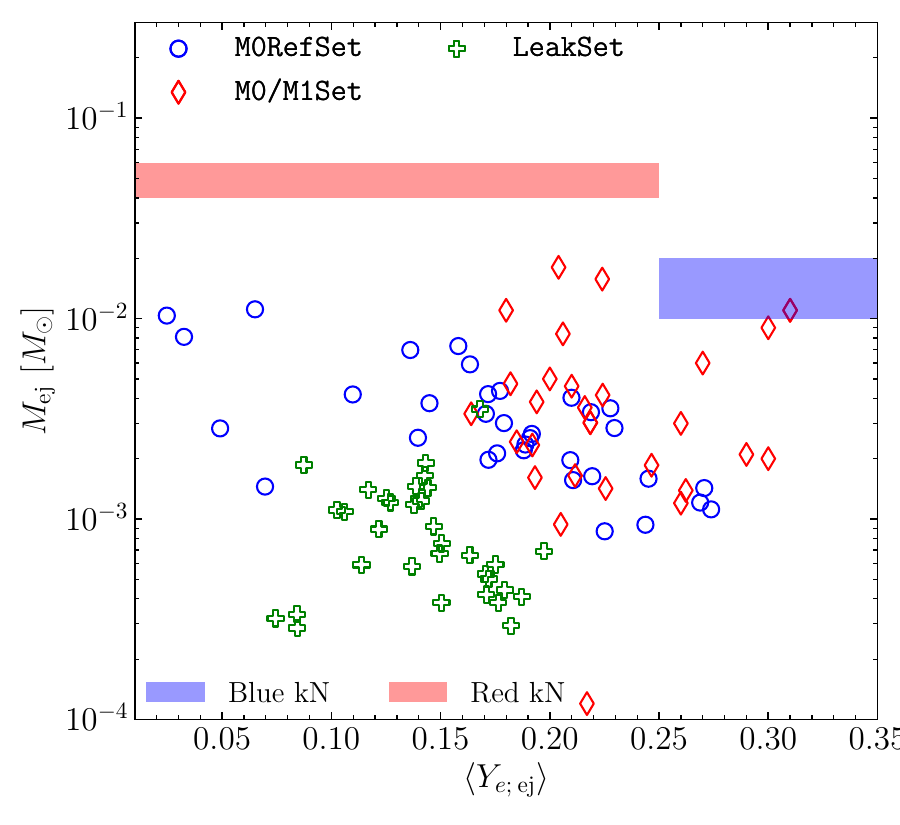}
    \includegraphics[width=0.32\textwidth]{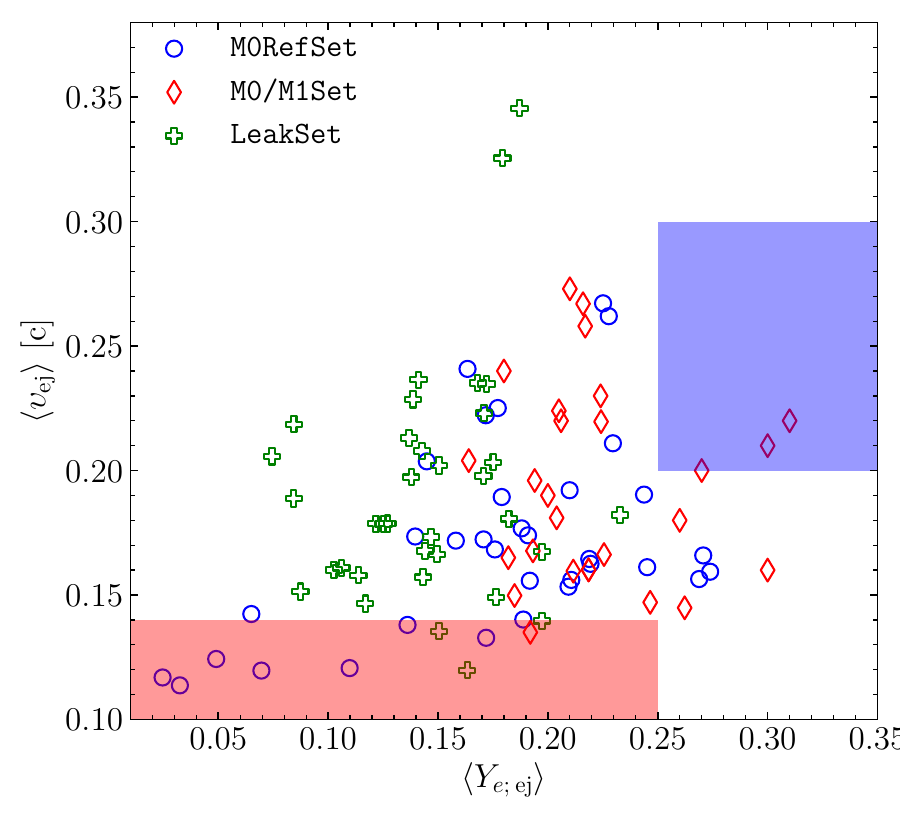}
    \caption{Summary of dynamical ejecta properties used in this work.
        Blue circles represent models of \DSrefset{}, 
        red diamonds stands for models from \DSheatcool{}, 
        green crosses are models from \DScool{}
        and gray squares stand for models from \DSnone{}, 
        We show for comparison the two-component fit to AT2017gfo as
        colored patches from \cite{Villar:2017wcc,Siegel:2019mlp}.
    }
    \label{fig:ejecta:dyn:ds}
\end{figure*}

In total we collect $324$ models. For $271$ of them we have/compute the 
binary parameters required for the analysis. 
For all of them the ejecta mass, $\amd$, is available.
For the models in \cite{Kiuchi:2019lls} the ejecta velocity is not reported, thus
only for $246$ models the mass-averaged ejecta velocity, $\avd$, is given.
In addition to \DSnone{} models, the average electron fraction of the ejecta is not provided
also in \cite{Lehner:2016lxy}. Hence, there are $99$ models for which the 
mass-averaged electron fraction of the dynamical ejecta, $\ayd$, is available.
Finally, for $76$ models the root mean square (RMS) half opening angle of 
the outflow about the equatorial plane, $\athetarms$, is available. 
The disk mass, $M_{\rm disk}$, is provided for $119$ models.

Since uncertainties estimates are not available for all
datasets, we assign errors following Ref.~\cite{Radice:2018pdn},
that were motivated by the observed resolution dependency of ejecta 
properties.
Different error measures, if adopted consistently, do not 
change results qualitatively, as we show in the case of \DSrefset{}
in Appendix~\ref{app:fiterror}.
For the dynamical ejecta mass we consider an uncertainty given by:
\begin{equation}
  \Delta M_{\text{ej}} = 0.5M_{\text{ej}} + 5\times10^{-5}M_{\odot}.
  \label{eq:ejecta:mej_err}
\end{equation}
For the ejecta velocity and for the electron fraction we consider 
$\Delta \upsilon_{\text{ej}} = 0.02$~c 
and $ \Delta Y_e = 0.01$ as fiducial uncertainties, respectively.
The latter value is justified by the robust behavior of the average electron 
fraction in simulations where multiple resolutions are available
\footnote{We expect larger uncertainties due to the approximate nature of current 
neutrino treatments (see \eg, \cite{Foucart:2016rxm,Foucart:2018gis}. 
However, due to the lack of extensive comparison studies, 
we consider only the numerical resolution error.}.

For the disk mass we assume \cite{Radice:2018pdn}
\begin{equation}
    \Delta M_{\text{disk}} = 0.5M_{\rm disk} + (5\times
    10^{-4})M_{\odot}\ .
    \label{eq:disk:mdisk_err}
\end{equation}

In this paper we aim to asses (i) the quality of the various 
fitting formulae to the ejecta properties and the disk mass. 
Because of the limited number of simulations in datasets, and 
having in mind multimessenger applications, 
instead of analyzing each dataset individually, in the main text 
we employ the following strategy. We study the progressively larger 
sample of simulations by iteratively adding datasets, starting from 
\DSrefset{}. The order in which we add the datasets is governed by the 
complexity of the physical setup, 
\ie, \DSheatcool{}, \DScool{} and finally \DSnone{}.
By progressively including datasets into the analysis we provide 
a suite of possible calibrations that can either contain the 
simulations with the most advanced physics input but relatively 
small number of them 
(\ie, \DSrefset{} and \DSheatcool{}), 
or all the simulations available. 
Using the standard statistical methods we rank the fitting 
formulae and discuss their application.
Additionally, we assess (ii) how the 
progressive inclusion changes the statistical properties of the 
enlarged set of simulations, aiming to assess 
the impact that simulating microphysics and neutrino transport has on the 
ejecta properties.
Finally, we elaborate on which fitting formula and what calibration 
are favorable based on our analysis in the discussion 
and directly apply it to modeling the key kilonova properties.

For (i) we consider the fitting formulae 
that exist in the literature as well as new fitting formulae based on
simple polynomials in the key BNS parameters \ie, 
reduced tidal deformability, $\tilde{\Lambda}$, and mass ratio, $q$.
Then we perform a standard fitting procedure with least square method, 
minimizing the residuals and display the fit performance 
on the residual plots for every quantity.
To quantitatively gauge the fit performance 
(for each ejecta property) we employ the 
sum of squared residuals (\ssr{}) defined as 
$\text{SSR} = \sum\limits_{i=1}^{N}(o_i - e_i)^2$ and the reduced $\chi^2$ statistics:
\begin{equation}
  \chi_{\nu}^{2} = \frac{\chi^2}{N - C} = \frac{1}{N-C}\sum\limits_{i=1}^{N}\Bigg(\frac{o_i - e_i}{o_i ^{\rm err}}\Bigg)^2,
  \label{eq:theory:chi2dof}
\end{equation}
where $N$ is the number of points in the dataset, $C$ is the number of coefficients in the 
fitting model (thus $N-C$ defines the number of degrees of freedom),
$o_i$ are the dataset values and $o_i ^{\rm err}$ their errors,
$e_i$ are the values predicted by the fitting model, and 
$o_i - e_i$ are the residuals.
The model comparison thus states that the 
lower \ssr{} is and the closer to
$1$ $\chi_{\nu}^{2}$ is, the better the model performs.
Note: a fit with the lowest $\chid$ may not necessarily be 
the fit with the lowest residuals if the error measure is not 
constant, \eg, for $\amd$ and $M_{\rm disk}$. 
This allows us to further asses the influence of the error measure.

This procedure is repeated for every dataset added. 
We provide the calibration for all fitting formulae and 
for all sets of datasets. 
We also perform the analysis for all datasets individually.
Results, reported in Appendix~\ref{app:datasets} corroborate 
the ones stated in the main text.

For (ii) we employ the following procedure. 
We start with the set that is uniform in physics and code, the \DSrefset{} 
that covers a narrow range in parameter space and allows to establish the base line. 
Then we add the rest of the models with neutrino heating and cooling effects, the \DSheatcool{},
and asses how the basic statistical properties have changed, employing 
the simplest quantitative measure that characterizes a statistical ensemble,
and standard deviation. 
To investigate the effects of the absence of neutrino reabsorption, we add the dataset 
that does not include this effect, the \DScool{} and repeat the analysis.
Finally, to asses the effect of the absence of neutrino cooling and differences in the EOS treatment 
we repeat the analysis with all datasets, including the \DSnone{}.
This iterative procedure allows to gauge 
the qualitative effect that different physical treatments 
have of the statistical behavior of the ejecta parameters and disk mass. 
We leave a more rigorous quantitative analysis to future works, 
when larger sample of data with physically motivated error measures 
and that cover a broader range in parameter space becomes available.

\section{Dynamical Ejecta}
\label{sec:dynej}

The mechanism behind the production of dynamical ejecta as well as the details 
on the numerical relativity simulations of \DSrefset{} 
are discussed in \textit{e.g.,} \cite{Radice:2020ddv,Bernuzzi:2020tgt,Nedora:2020pak}. 
Here, we focus on overall properties of the mass ejecta in 
relation to other results in the literature, and provide approximate 
fitting formulae for the total ejecta mass, 
the mass-averaged velocity, the electron fraction and the RMS half opening angle. 
Importantly, the are several criteria for a fluid element to 
become gravitationally unbound -- to become ejecta. Due to the 
"burst-like" nature of dynamical ejecta, the geodesic criterion, 
that considers fluid elements moving on ballistic trajectories, 
neglecting the fluid pressure 
\cite{Rezzolla:2013}, is commonly employed \cite{Radice:2016dwd,Radice:2018pdn,Bernuzzi:2020txg}. 
Another broadly used criterion is the Bernoulli criterion, that 
takes into account the internal energy of the fluid. 
With respect to the dynamical ejecta, these two criteria was found to lead 
to the ejecta mass estimations different by a factor of $2$ \cite{Foucart:2015gaa}.
Additionally, depending on the length of the postmerger evolution 
of a simulation, different methods are employed to compute the ejecta properties.
For instance, in \cite{Radice:2018pdn}, the ejecta was computed following 
the matter passing an extraction sphere untill the matter flux subsided. 
Simulations were sufficiently long to allow the mass flux to saturate. 
Meanwhile in \cite{Vincent:2019kor}, a combination ejecta that was able 
to leave the simulation domain and that was still within the domain of the 
simulation at the end was considered. 
These differences in ejecta criteria and method of calculation 
present an additional source of systematics in data.

Figure \ref{fig:ejecta:dyn:ds} summarizes the total mass, the mass-averaged
velocity and mass-averaged electron fraction (where available) for the used datasets. 
Overall we note that the ejecta properties of the models of \DSrefset{} are compatible 
with those of \DSheatcool{}, as they include the same physics with respect
to the EOS treatment and also include the effect of neutrino absorption. 
Notably, the very high mass-ratio, $q$, models of \DSrefset{},
discussed in \cite{Bernuzzi:2020txg}, show slightly different properties,
as their ejecta is of tidal origin only.
Comparing the properties of \DSheatcool{} and \DScool{} we observe that 
neutrino absorption leads, on average, to a larger ejecta mass, which is 
especially noticeable for the leakage subset of \cite{Radice:2018pdn}(LK).
Additionally, neutrino absorption leads to a higher $\ayd$ of the ejecta, while
the average velocity, $\avd$, appears to be independent of it.

In the following we discuss the fitting formulae for the different quantities.

\subsection{Mass}

In order to asses the systematic changes in ejecta masses between different 
datasets with different physics input, we restrict the binary parameter space 
to $q\in(1,1.2)$ and $\tilde{\Lambda}\in(350,850)$, common for all datasets 
that we compare. In doing so we reduce the number of simulations significantly. Thus, we aim to assess the changes on the 
qualitative level only. A more rigorous analysis 
would require significantly larger sample of simulations, homogeneously distributed in the parameter space.
The dynamical ejecta mass, averaged over $8$ simulations of \DSrefset{} is
$\overline{\amd} = (3.5 \pm 1.3)\times 10^{-3}M_{\odot}\,$ 
where hereafter we report also the standard deviation computed over the
relevant simulation sample~\footnote{
    We report here the mean value as it
    is the simplest quantitative measure to characterizes the
    differences between the different datasets.
}.

Adding the rest of \DSheatcool{} 
(7 models)
we obtain $(5.1 \pm 3.9)\times 10^{-3}M_{\odot}$.
The increase is 
given largely by datasets that include the M1 neutrino scheme, 
\cite{Vincent:2019kor} and \cite{Sekiguchi:2016bjd}. 
However, adding models of \DScool{}, 
(another 8 models) 
we observe that the mean
value decreases to $3.8\times 10^{-3}M_{\odot}$, as 
models without neutrino absorption predict, on average,
lower ejecta masses.
Finally, adding models without neutrinos at all, some of 
which have polytropic EOS 
(7 models in the restricted parameter space), we do not 
observe change in the $\overline{\amd}$.

Lifting the restrictions on the parameter space, 
we fit all the available data 
using second-order polynomials in one parameter
$(\tilde\Lambda)$, and in two parameters, $(q,\tilde\Lambda)$, namely:
\begin{align}\label{eq:polyfit2}
  P_2 ^1(\tilde{\Lambda}) &= b_0 + b_1\tilde\Lambda + b_2 \tilde\Lambda^2, \\\label{eq:polyfit22}
  P_2 ^2(q,\tilde\Lambda) &= b_0 + b_1q + b_2\tilde\Lambda + b_3q ^2 +  b_4 q \tilde\Lambda + b_5\tilde\Lambda^2 \, .
\end{align}
Additionally, we consider the fitting model presented
Refs.~\cite{Kawaguchi:2016ana,Dietrich:2016fpt,Radice:2018pdn}
\begin{align}
  \label{eq:fit_Mej}
  &\left(\frac{\amd}{10^{-3}M_{\odot}}\right)_{\rm fit} =
  \Big[\alpha\Big(\frac{M_B}{M_A}\Big)^{1/3}\Big(\frac{1-2C_A}{C_A}\Big)+  
    \beta\Big(\frac{M_B}{M_A}\Big)^n \\
    &+ \gamma\Big(1-\frac{M_A}{M_{b\,A}} \Big)\Big]M_{b\,A} + (A\leftrightarrow B) + \delta\non,
\end{align}
and the model presented in \cite{Kruger:2020gig}:
\begin{equation}
  \label{eq:fit_Mej_Kruger}
  \left(\frac{\amd}{10^{-3}M_{\odot}}\right)_{\rm fit} =
  \left(\frac{\alpha}{C_A} + \beta\frac{M_B ^n}{M_A ^n} + \gamma
  C_A\Bigg)M_A + (A\leftrightarrow B)\right. \ .
\end{equation}

As in some cases the values of $\md$ change by orders of magnitude for
very close values of $q$ and $\tilde{\Lambda}$, we  
calibrate the fitting models to $\log_{10}(\md)$ instead of the $\md$.

Regarding Eq.~\eqref{eq:fit_Mej} and Eq.~\eqref{eq:fit_Mej_Kruger},
we also note that these formulae deliver ill-conditioned fits, with
coefficients that change up to a factor of two for the same data,
depending on the guesses or on the nonlinear fitting algorithm employed. 
While such formulae may allow to account for a non-smooth behavior 
in data, their calibration presents an additional challenge.

Fitting coefficients as well as values of $\chid$ 
are reported in Appendix~\ref{app:coefs}:
coefficients of the polynomial regressions are reported in 
Tab.~\ref{tab:dynfit:poly};
fits coefficients for Eqs.\eqref{eq:fit_Mej}-\eqref{eq:fit_Mej_Kruger}
are reported in Tab.~\ref{tab:dynfit:fit_form}.


\begin{table}[t]
    \caption{
        Values of \ssr{} for different 
        fitting models for the dynamical ejecta properties. Mean is the simulation
        average, $P_n(x,y)$ is a polynomial of order $n$ in the variables $x,y$. Fits are performed for the data of this work and for an increasingly larger combined dataset from
        the literature. See text for discussion. 
        The best fitting model for a given dataset is characterized by the 
        lowest value of \ssr{}.
    }
    \label{tbl:fit:ejecta:chi2dofsall}
    \scalebox{0.88}{
        \begin{tabular}{l|l|ccccc}
            \hline\hline
            $\log_{10}(\md)$ & Datasets & Mean & Eq.~\eqref{eq:fit_Mej} & Eq.~\eqref{eq:fit_Mej_Kruger} & $P_2^1(\tilde{\Lambda})$ & $P_2^2(q,\tilde{\Lambda})$ \\ \hline
            & \DSrefset{} & 2.57 & 1.65 & 1.40 & 2.43 & 0.97 \\ 
            & \& \DSheatcool{} & 8.19 & 7.51 & 6.35 & 7.84 & 6.55 \\ 
            & \& \DScool{} & 33.13 & 26.37 & 21.57 & 29.62 & 24.40 \\ 
            & \& \DSnone{} & 86.93 & 80.08 & 63.38 & 86.85 & 55.09 \\ 
            \hline\hline
            $\langle v_{\rm ej}\rangle$ & Datasets & Mean & Eq.~\eqref{eq:fit_vej} & & $P_2^1(\tilde{\Lambda})$ & $P_2^2(q,\tilde{\Lambda})$ \\ \hline
            & \DSrefset{} & 0.04 & 0.02 & & 0.04 & 0.01 \\ 
            & \& \DSheatcool{}  & 0.09 & 0.05 & & 0.07 & 0.04 \\ 
            & \& \DScool{}  & 0.29 & 0.24 & & 0.25 & 0.21 \\ 
            & \& \DSnone{}  & 0.78 & 0.66 & & 0.74 & 0.67 \\ 
            \hline\hline
            $\langle Y_{\rm e}\rangle$ & datasets & Mean & & & $P_2^1(\tilde{\Lambda})$ & $P_2^2(q,\tilde{\Lambda})$ \\ \hline
            & \DSrefset{} & 0.14 & & & 0.13 & 0.02 \\
            & \& \DSheatcool{}  & 0.24 & & & 0.23 & 0.06 \\ 
            & \& \DScool{}  & 0.35 & & & 0.35 & 0.23 \\ 
            \hline\hline
            $\langle \theta_{\rm RMS}\rangle$ & datasets & Mean & & & $P_2^1(\tilde{\Lambda})$ & $P_2^2(q,\tilde{\Lambda})$ \\ \hline
            & \DSrefset{} & 2775 & & & 2631 & 498 \\ 
            & \& \DSheatcool{}  & 2949 & & & 2788 & 574 \\ 
            & \& \DScool{}  & 4681 & & & 4116 & 2382 \\ 
            \hline\hline
        \end{tabular}
    }
\end{table}

Different fits for the dynamical ejecta properties are compared in
terms of the sum of squared residuals, \ssr, in Tab.~\ref{tbl:fit:ejecta:chi2dofsall}.
We find that fitting the data from only \DSrefset{} as well as all the data from 
all datasets combined, the lowest \ssr{} is given by \polql{}. 
The Eq.~\eqref{eq:fit_Mej_Kruger} gives similar, albeit slightly larger values for 
these sets of simulations, while performing slightly better for the other two combinations 
of datasets. 
Invoking the error measure and the $\chid{}$ statistic we observe a very similar 
picture with \polql{} giving the lowest $\chid$ when all datasets are considered 
and Eq.~\eqref{eq:fit_Mej_Kruger} performing better when only \DSheatcool{} 
and \DSrefset{} are considered. 
The small difference in performance 
between these two fitting formulae 
can be attributed to the fact that both include the mass 
ratio explicitly, which allows to capture the leading trend in the data. 

\begin{figure}[t]
    \centering 
    \includegraphics[width=0.49\textwidth]{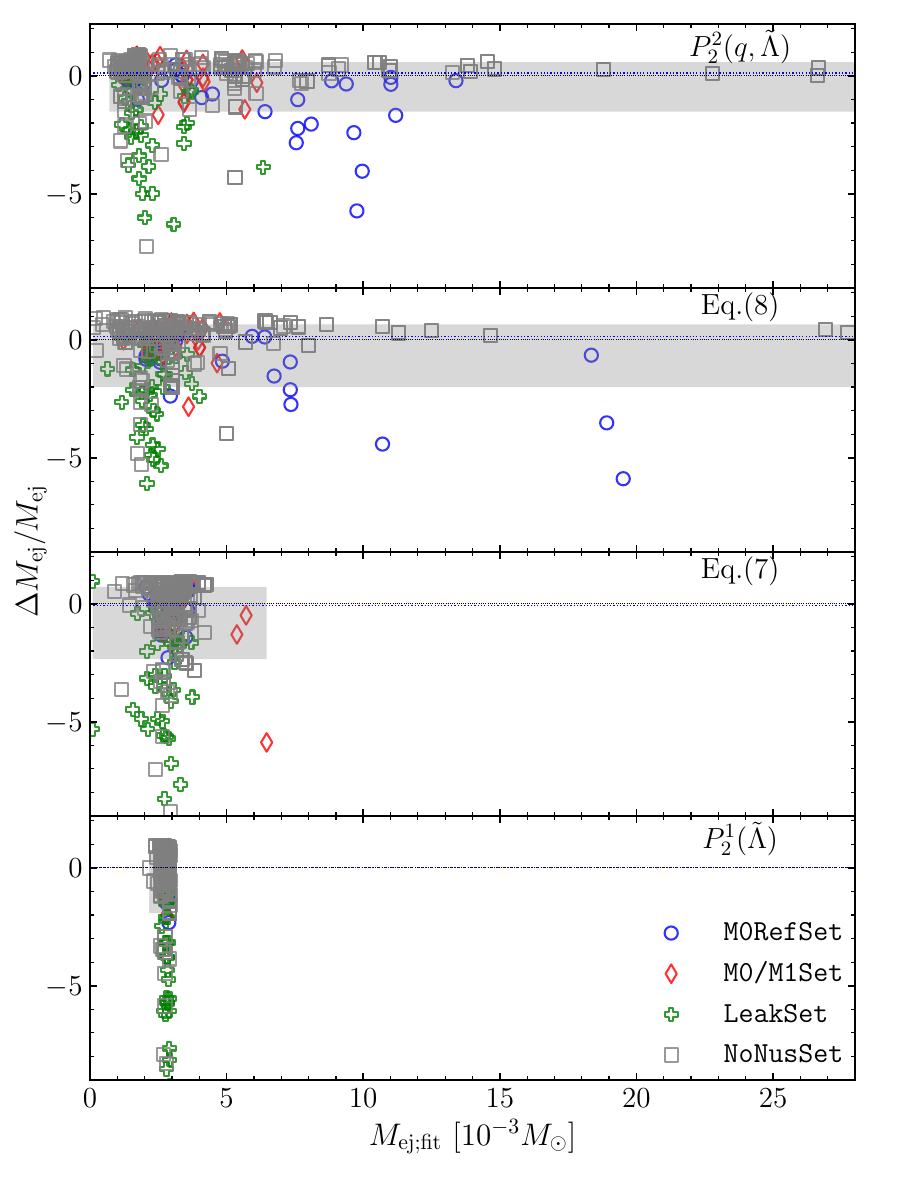}
    \caption{
      Relative differences between data and fits for the dynamical ejecta
      mass, $\Delta M_{\rm ej} = M_{\rm ej} - M^{\rm fit}_{\rm ej}$.
      We show polynomial fits and fitting formulae
      Eq.~\eqref{eq:fit_Mej} and Eq.~\eqref{eq:fit_Mej_Kruger} 
      calibrated with all datasets available.
      From top to bottom the models arrange based on their 
      \ssr: from lowest to highest
      See Tab.~\ref{tbl:fit:ejecta:chi2dofsall}. 
      The gray region represents the fit's $68\%$ confidence level.
      Note that fitting was performed minimizing $\log_{10}(\md)$. See text for details.
    }
    \label{fig:ejecta:dyn:m}
\end{figure}

The Eq.~\eqref{eq:fit_Mej} cannot sufficiently well reproduce the low 
ejecta masses of models with microphysic EOS and leakage neutrino 
transport scheme and high ejecta masses of models with polytropic 
EOS and no neutrino transport. This results in the truncated 
$M_{\rm ej;fit}$ (see Fig.~\ref{fig:ejecta:dyn:m}) and larger
\ssr{} and $\chid$.  
In addition, it was previously found in \cite{Radice:2018pdn}, 
that this fit model does not accurately reproduce the data
and misses systematic trends.
The simplest model, a polynomial of only $\tilde{\Lambda}$,
cannot capture leading trends in the data, resulting in 
considerably larger \ssr{} and $\chid{}$
once all datasets are considered. Similarly, taking the simple mean value 
as a fitting model results in an even larger 
\ssr{} and $\chid$.
Thus, the inclusion of the dependency on mass-ratio is of 
crucial importance for modeling dynamical ejecta mass.

In Fig.~\ref{fig:ejecta:dyn:m} we show the relative differences
between all datasets values and values from the fitting models. 
We observe that none of the fitting models can
adequately capture the subset of \DScool{} 
with a leakage scheme only as neutrino treatment,
(Cf. \cite{Radice:2018pdn,Lehner:2016lxy}).
While the lowest \ssr{} and $\chid{}$ are found for \polql{},
the plot shows that the Eq.~\eqref{eq:fit_Mej_Kruger} can also
capture the large ejecta mass of \DSnone{} and \DSrefset{} 
with however higher residuals. 
Notably Eq.~\eqref{eq:fit_Mej} cannot capture that tail,
truncating the distribution at $\sim10^{-2}M_{\odot}$. 
The polynomial in $\tilde{\Lambda}$ fits the data very poorly,
showing an almost flat distribution around the mean value of 
the ejecta mass.

\subsection{Mass-averaged velocity}

\begin{figure}[t]
    \centering 
    \includegraphics[width=0.49\textwidth]{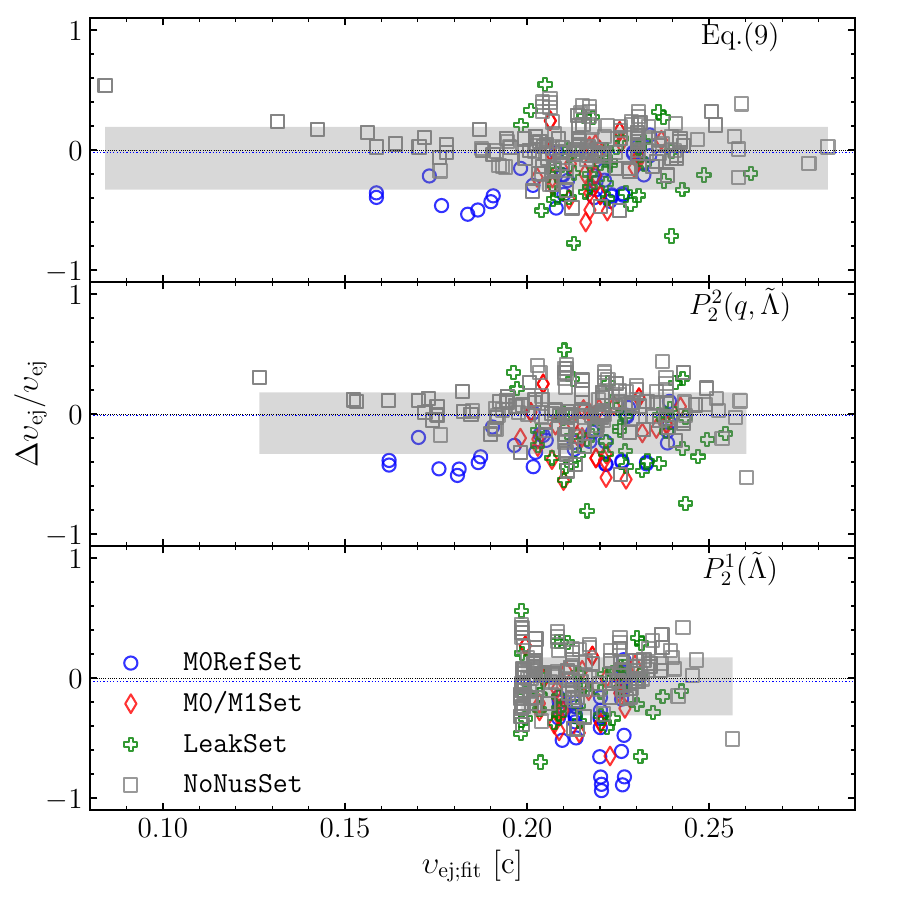}
    \caption{
      Relative differences between data and fits for the mass-averaged velocity of the dynamical ejecta, $\Delta \upsilon_{\rm ej} = \upsilon_{\rm ej} - \upsilon^{\rm fit}_{\rm ej}$.
      Calibration is done for all datasets available. 
      We show the fitting formula Eq.~\eqref{eq:fit_vej} and the polynomial fits.
      From top to bottom the models are arranged based on their $\chid{}$: from lowest to highest.
    }
    \label{fig:ejecta:dyn:v}
\end{figure}

\begin{figure}[t]
    \centering 
    \includegraphics[width=0.49\textwidth]{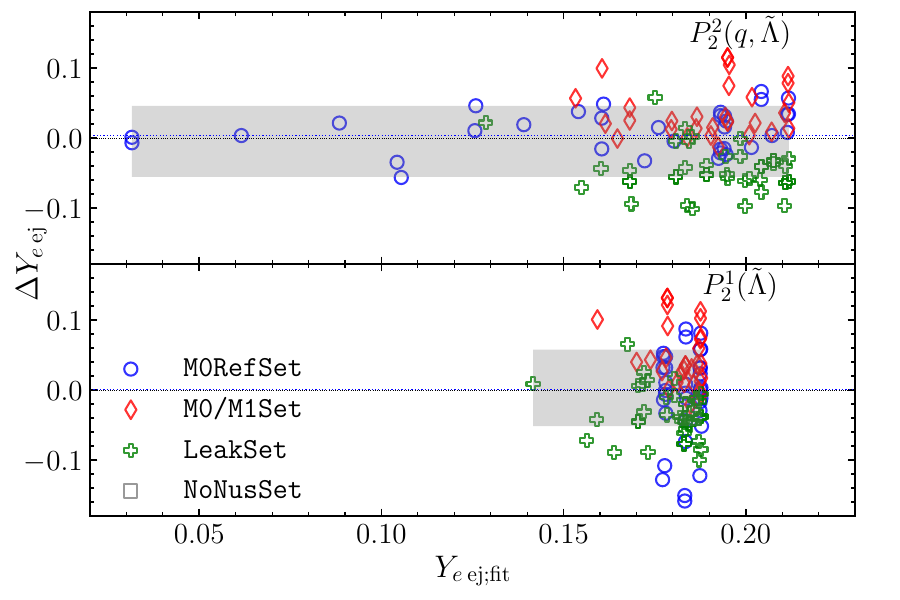}
    \caption{
        Relative differences between data and fits for the 
        mass-averaged electron fraction of the dynamical ejecta.
        We show the polynomial fits, and Eq.~\eqref{eq:polyfit2} and Eq.~\eqref{eq:polyfit22}.
        Calibration is done for all datasets available. 
        Here $\Delta Y_{e\: \rm ej} = Y_{e\: \rm ej} - Y^{\rm fit}_{e\: \rm ej}$.
        From top to bottom the models are arranged based on their $\chid{}$: from lowest to highest.
    }
    \label{fig:ejecta:dyn:y}
\end{figure}

The mass-averaged terminal velocity of the dynamical ejecta, $\avd$, from \DSrefset{}
ranges from $0.11\, c$ to $0.27\,c$, 
in agreement with the leakage simulations performed with the same code in 
\cite{Radice:2018pdn}.
However, differently from the analysis of \cite{Radice:2018pdn}, 
the correlation of the $\langle\upsilon_{\infty}\rangle$ with the tidal
parameter $\tilde{\Lambda}$ was found in the models of 
\DSrefset{} with the fixed chirp mass \cite{Nedora:2020pak}.
Models with lower $\tilde{\Lambda}$, showed higher velocities.
This is a consequence of the fact that the dynamical ejecta in comparable-mass 
mergers are dominated by the shocked component and that 
the shock velocity is larger the more compact the binary is.
On the contrary, for high mass ratios $q\gtrsim1.5$, the ejecta are
dominated by the tidal component and a larger $q$ leads to a smaller
$\avd$ in \DSrefset{}.

Restricting the parameter space again, we asses the change in mean value 
of ejecta velocity, $\overline{\avd}$.
For the models of \DSrefset{} we find 
$\overline{\avd}=0.19 \pm 0.03\,c$. 
When we iteratively add models of \DSheatcool{}, \DScool{} and \DSnone{}, the 
$\overline{\avd}$ remains largely unchanged, taking values of 
$0.20\,c$, $0.20\,c$ and $0.21\,c$. 
Notably, Fig.~\ref{fig:ejecta:dyn:ds}, shows that some models of \DSnone{} (models of Cf.~\cite{Bauswein:2013yna}
\footnote{
In \cite{Bauswein:2013yna} the different treatment of gravity was employed. 
Specifically, the evolution was performed under the assumption of conformal flatness.
}) 
have an overall larger velocity.
However, they lie outside of the restricted parameter space.

Lifting the restrictions on the parameter space 
we fit the data with a second-order polynomials, as in 
Eq.~\eqref{eq:polyfit2}, and also with the fit formula reported in 
\cite{Dietrich:2016hky,Radice:2018pdn}:
\begin{equation}
{\avd}_{\rm fit} = \Big[\alpha\Big(\frac{M_A}{M_B}\Big)(1+\gamma C_A)\Big] + (A\leftrightarrow B) + \beta
\label{eq:fit_vej} \, .
\end{equation}
We note, that similarly to the Eq.~\eqref{eq:fit_Mej} and Eq.~\eqref{eq:fit_Mej_Kruger},
the outcome of the calibration of the Eq.~\eqref{eq:fit_vej} depends on the 
initial guesses of the minimization algorithm.

The coefficients of the polynomial regressions for $\avd$ are reported in
Tab.~\ref{tab:dynfit:poly};
fits coefficients for Eq.~\eqref{eq:fit_vej} are reported in 
Tab.~\ref{tab:dynfit:fit_form}.
The fit models' performance is summarized in Tab.~\ref{tbl:fit:ejecta:chi2dofsall}
in terms of \ssr{}.

We find that unless models of \DSnone{} are included, the \polql{} 
displays the lowest \ssr{} and $\chid{}$ among other fitting formulae. 
When models of \DSnone{} are also included, the \polql{} and 
Eq.~\eqref{eq:fit_vej} perform rather similar.

\begin{figure*}[t]
    \centering 
    \includegraphics[width=0.32\textwidth]{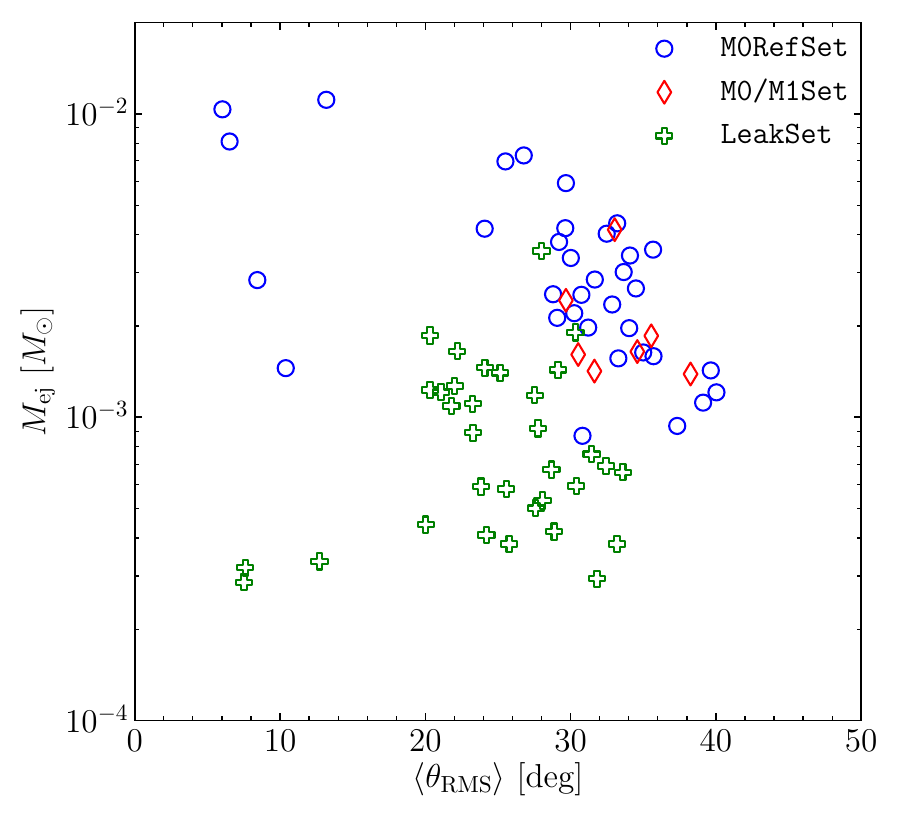}
    \includegraphics[width=0.32\textwidth]{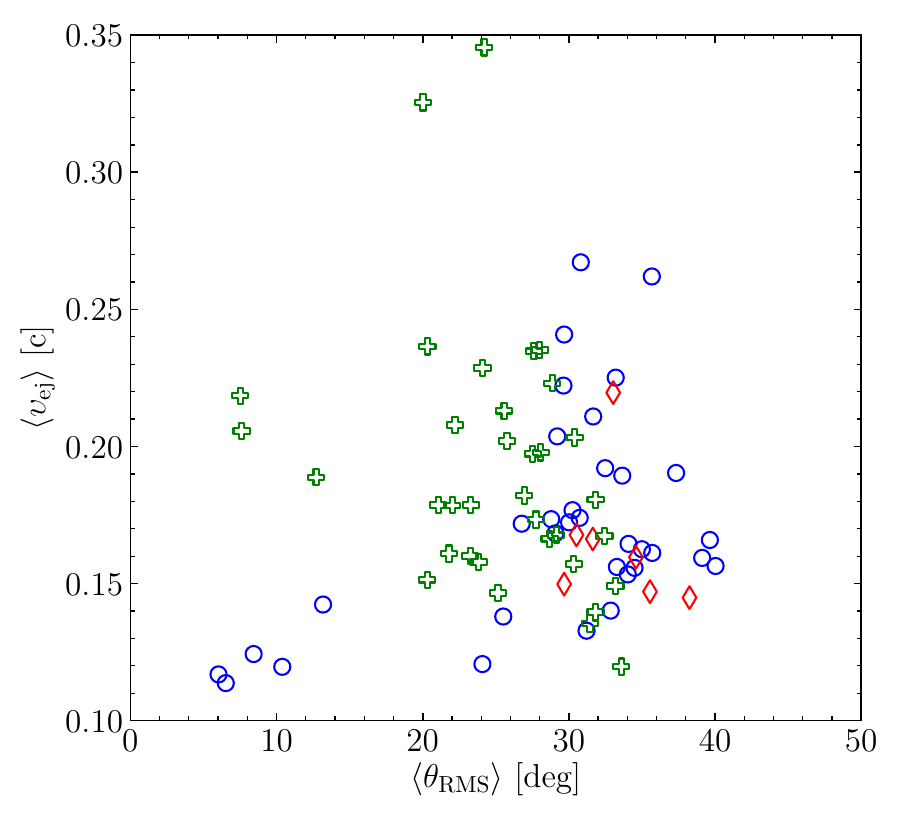}
    \includegraphics[width=0.32\textwidth]{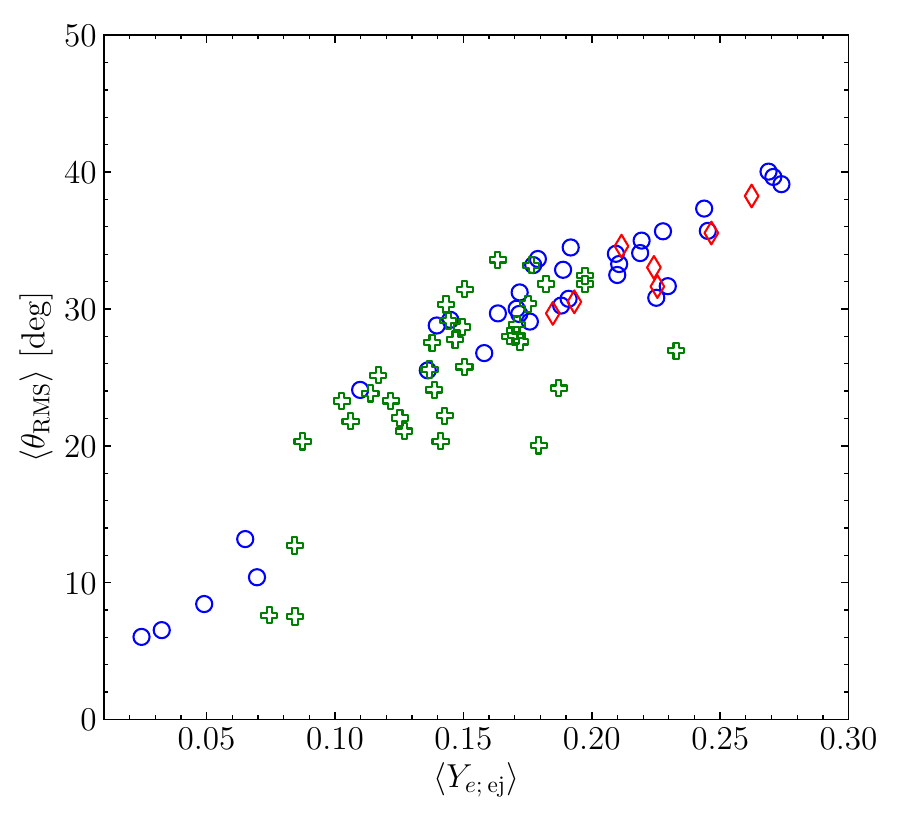}
    \caption{
        Relations between the ejecta $\theta_{\rm RMS}$ and other parameters of the dynamical
        ejecta: mass, $\amd$, velocity, $\avd$, and electron fraction $\ayd$ for models from
        \DSrefset{} and \cite{Radice:2018pdn} from \DScool{} and \DSheatcool{}.
        Plots show that models with neutrino absorption have
        higher $\amd$ and larger $\theta_{\rm RMS}$ as well as 
        a clear correlation between $\theta_{\rm RMS}$ and $\ayd$.
    }
    \label{fig:ejecta:dynej_thetarms}
\end{figure*}

\begin{figure}[t]
    \centering 
    \includegraphics[width=0.49\textwidth]{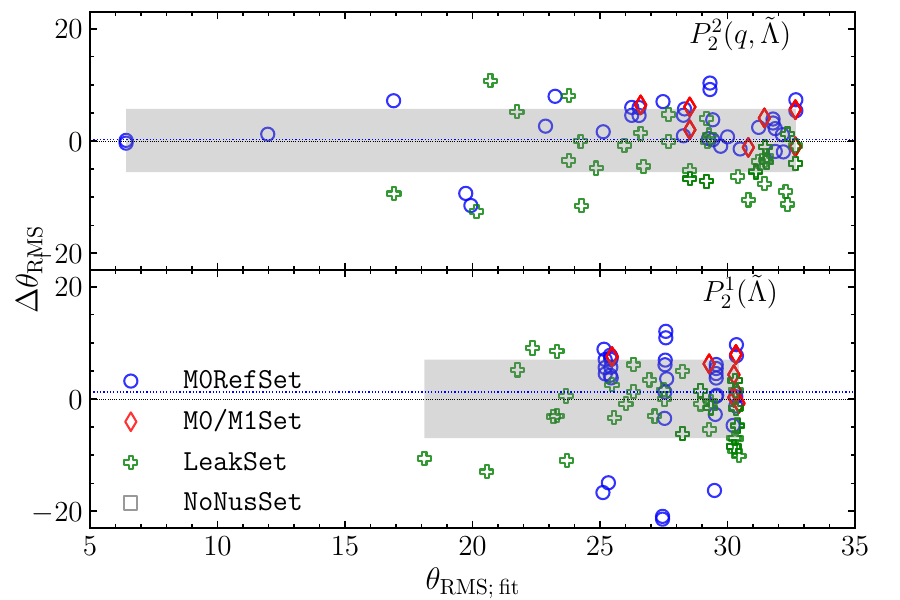}
    \caption{
        Relative differences between data and fits of dynamical
        ejecta mass-averaged electron fraction.
        We show polynomial fits only.
        Here $\Delta \theta_{\rm RMS} = \theta_{\rm RMS} - \theta^{\rm fit}_{\rm RMS}$.
        From top to bottom the models are arranged based on their $\chid$: from lowest to highest.
    }
    \label{fig:ejecta:dyn:theta}
\end{figure}

In Fig.~\ref{fig:ejecta:dyn:v} we show the differences 
between the data and the fits for the considered fitting models. 
We find that Eq.~\eqref{eq:fit_vej} and the second
order polynomial in ($q,\tilde{\Lambda}$) reproduce most of the data
within an error of $\sim50\%$ and overall perform very similarly. 
In both cases, the largest deviations are obtained for models 
of the \DScool{}, with the neutrino leakage scheme. 
The one parameter polynomial of $\tilde{\Lambda}$
fails to capture the low velocity tail of the 
distribution and overall gives considerably higher 
differences between the dataset and the model predicted values of 
$\avd$.

\subsection{Electron fraction} 

The mass-averaged electron fraction, $\ayd$, in \DSrefset{}
varies from $0.03$ to $0.27$.

Restricting the parameter space to the common region, we obtain the mean value of 
electron fraction for \DSrefset{}  $\overline{\ayd}=0.19\pm0.02$. 
Adding models of \DSheatcool{} increases the mean to $0.20\pm0.04$ which 
is largely due to models of \cite{Sekiguchi:2015dma,Sekiguchi:2016bjd}
with leakage+M1 scheme (see Fig.~\ref{fig:ejecta:dyn:ds}).
When models of the \DScool{} are added, the mean values decreases back, 
which is as expected as models with leakage scheme only have lower 
ejecta electron fraction \eg, \cite{Radice:2018pdn}. 
Notably, the number of simulations added is rather small 
\footnote{Note that \cite{Lehner:2016lxy} does not provide electron fraction.}.

Regarding the fitting functions, we explore the low-order polynomials 
in $(q,\tilde\Lambda)$ and in $(\tilde\Lambda)$ only.
The coefficients of polynomial regressions are reported in Tab.~\ref{tab:dynfit:poly}.
We observe that for all datasets, the \polql{} displays consistently 
lower \ssr{} and $\chid{}$. Notably, the addition of \DScool{} models 
leads to a jump in these measures, as the data in this set is statistically 
different (different physics setup).
In Fig.~\ref{fig:ejecta:dyn:y} we show the performance of the different 
fitting models for the mass-averaged electron fraction of the ejecta. 
When all datasets are considered, the second order polynomial manages to 
reproduce both the low-$Y_e$ tail and high $Y_e$ values for models
with advanced neutrino treatment.
The accurate computation of the electron fraction naturally 
requires neutrino absorption to be included into simulation setups. 
The availability of a larger number of simulations with advanced 
neutrino transport will undoubtedly improve fitting models.

\subsection{Root mean square half opening angle} 

Ejecta geometry was found to have a strong imprint on the properties of the 
electromagnetic counterparts to mergers \citep[\eg][]{Perego:2017wtu}. 
Numerical relativity simulations show that the form of the angular distribution 
of ejecta properties is quite complex \citep[\eg][]{Radice:2018pdn} and 
presents challenges for a statistical analysis. 
Here we employ the mass-averaged RMS half opening angle 
(under the assumption of axial symmetry), 
a quantity that can be used to separate the massive, low-latitude 
outflow and less massive, polar one.
In the Discussion we show an example of how this quantity can be used in 
kilonova modeling. 
Following \cite{Radice:2018pdn}, we define the 
mass-averaged RMS half opening angle as by assuming axial symmetry and computing:
\begin{equation}
    \theta_{\rm RMS} = \frac{180}{\pi}\Bigg(\frac{\sum m_i \theta_i^2}{\sum m_i}\Bigg)^{1/2}\, ,
\end{equation}
where $\theta_i$ and $m_i$ are the angle (from the binary plane) 
and mass of the ejecta element. 
This quantity is available only for \DSrefset{} and for the models of 
\cite{Radice:2018pdn}.
In Figure~\ref{fig:ejecta:dynej_thetarms} we show the dependency of $\theta_{\rm RMS}$
on the previously discussed ejecta parameters. 
Comparing the data from \DSrefset{} and the leakage dataset of Ref.~\cite{Radice:2018pdn}, 
we find that the inclusion of neutrino absorption leads to larger 
$\theta_{\rm RMS}$ on average with the exception of highly asymmetric models of \DSrefset{}.
Notably we observe a clear linear relation between the $\theta_{\rm RMS}$ and $\ayd$ 
(see Fig.~\ref{fig:ejecta:dynej_thetarms}).
The origin of this relation lies in the dependency of the ejecta properties on the binary mass-ratio. 
Asymmetric binaries produce low-$Y_e$, tidal ejecta confined largely to the lane of the binary, 
while for more symmetric models with
prominent shocked ejecta component there is a trend to have higher $Y_e$ and more spread-out ejecta. 
This further suggests that $\theta_{\rm RMS}$ can help capturing the 
transition between the low- and high-opacity ejecta in kilonova modeling. 

The number of models within the restricted parameter space for which we 
have the $\theta_{\rm RMS}$ is very limited. Thus we only report the 
average value for \DSrefset{}, 
$\overline{\athetarms}=(31.7 \pm 1.9)~\text{deg}$.

In light of the considerably smaller sample of models for which we have
$\theta_{\rm RMS}$, we simplify the statistical analysis, 
considering as fitting models only polynomials: 
$P_2^1(\tilde{\Lambda})$ and \polql{}.
The coefficients of the polynomial regressions are reported in
Tab.~\ref{tab:dynfit:poly}.
Following \cite{Radice:2018pdn} we adopt a uniform error for all models 
of $2$ degrees.
We find that similarly to the case of ejecta electron fraction, the 
\polql{} performs consistently better here than
other options for all datasets in terms of both \ssr{} and $\chid$.

In Fig.~\ref{fig:ejecta:dyn:theta} we show the performance of polynomial 
fitting models to the ejecta $\theta_{\rm RMS}$. 
The second order polynomial provides a better fit to 
the low-$\theta_{\rm RMS}$ tail of the distribution than $P_2^1(\tilde{\Lambda})$.
and reproduces the data within $\sim 10\,$deg. 
Overall, we observe that the inclusion of both $q$ and $\tilde{\Lambda}$ 
in a fitting formula is important for capturing the trends in data.
However, the small sample of models does 
not allow us to conduct a more thorough investigation, in particular,
to study the effects of various physics included in simulations. 

\begin{figure*}[t]
    \centering 
    \includegraphics[width=0.49\textwidth]{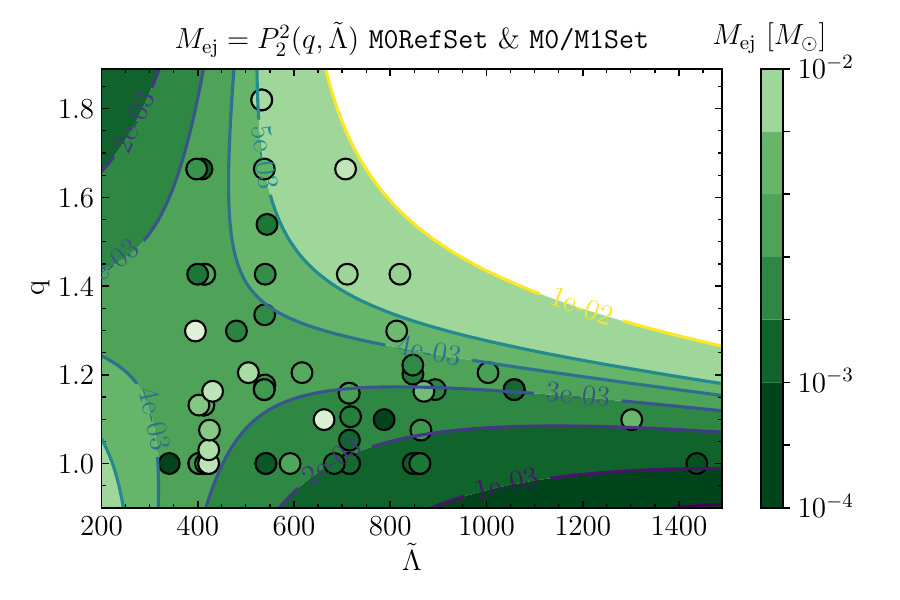}
    \includegraphics[width=0.49\textwidth]{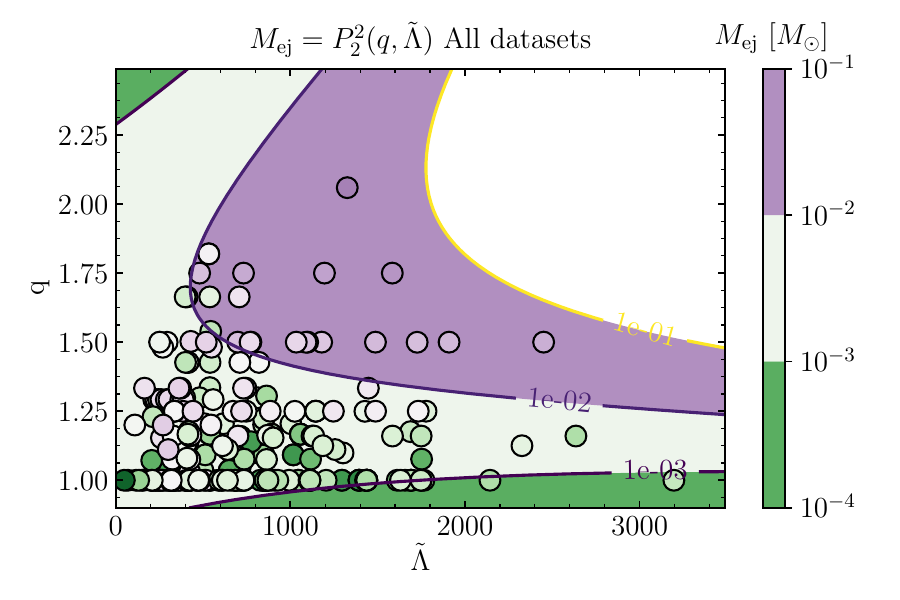}
    \includegraphics[width=0.49\textwidth]{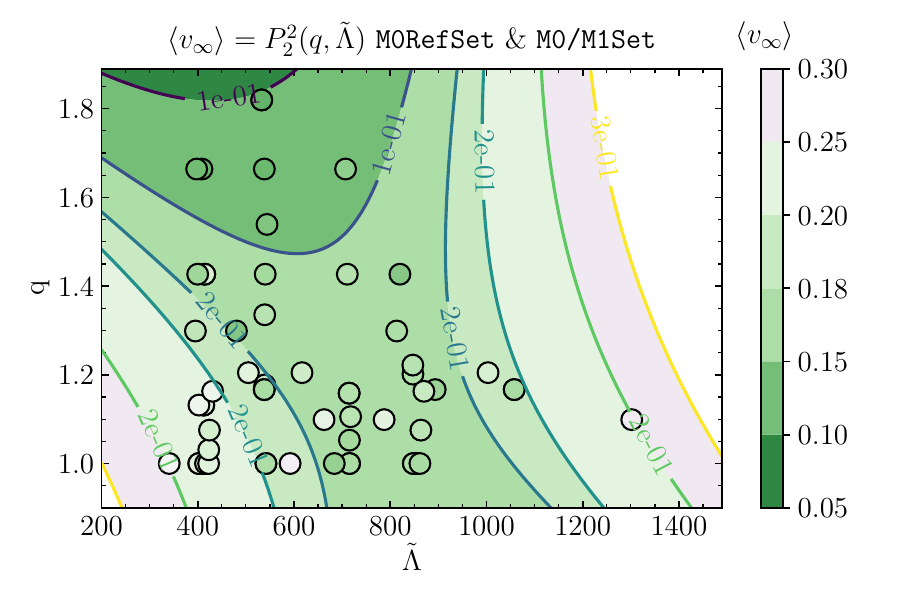}
    \includegraphics[width=0.49\textwidth]{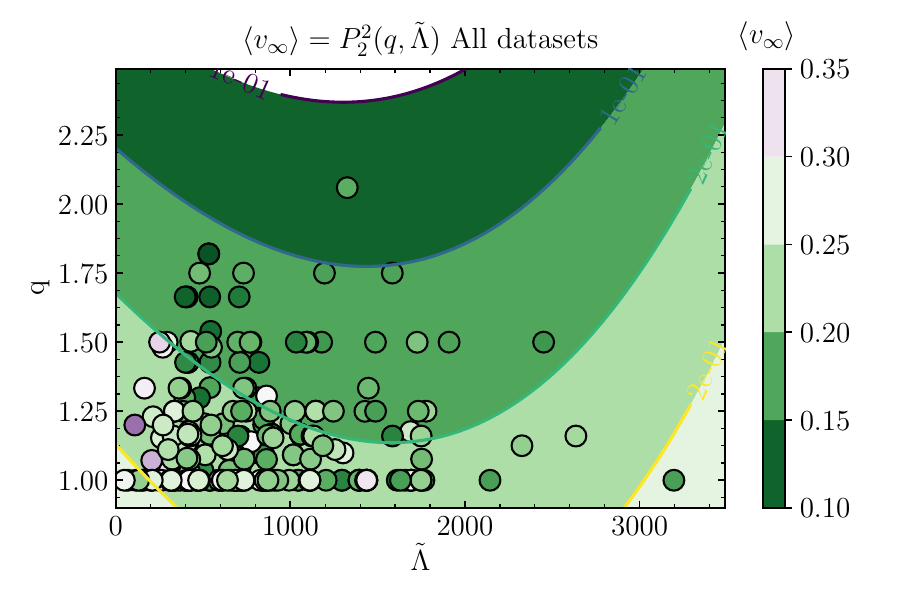}
    \includegraphics[width=0.49\textwidth]{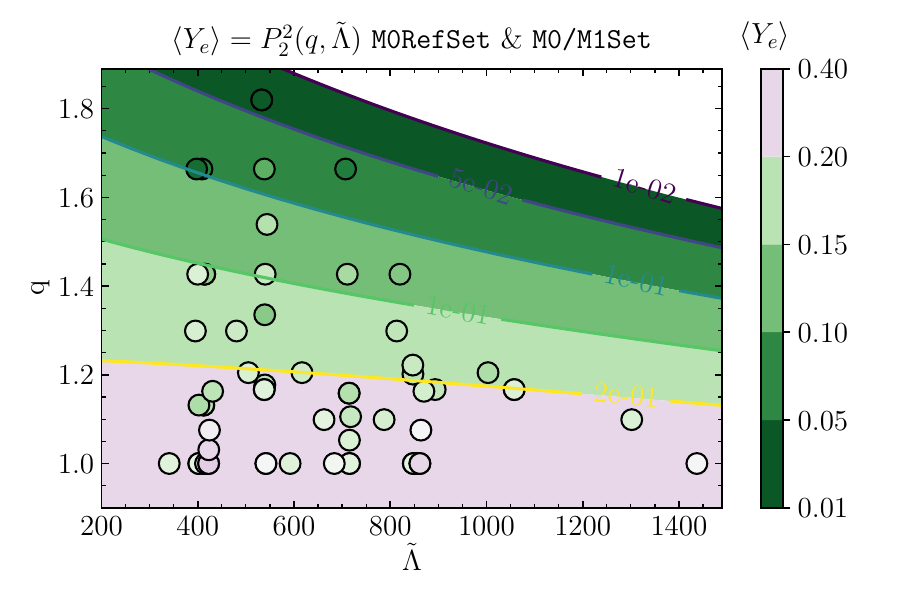}
    \includegraphics[width=0.49\textwidth]{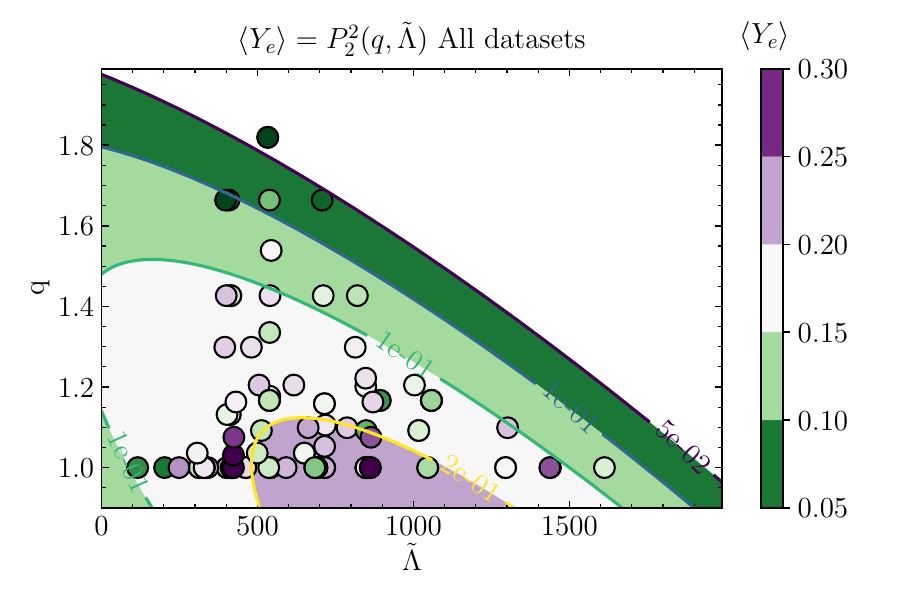}
    \caption{
        Comparsion between ejecta parameters informed by the fit 
        (colored contours), 
        and the simulation ejecta data (colored markers) for \polql{}  
        fitting model calibrated with advanced-physics datasets, 
        \DSrefset{} and \DSheatcool{}, (\emph{left column of panels}) and with all 
        available datasets (\emph{right column of panels}).
        The plot shows that for some physical quantities, such as ejecta 
        electron fraction and velocity, the leading trends in data appear to be 
        captured by the \polql{} calibrated with datasets with advanced physics.
        When all datasets are considered, however, the limitations of the 
        smooth polynomial fitting function becomes apparent as it is not 
        able to fit the non-smooth data well.
    }
    \label{fig:ej_parspace}
\end{figure*}

\subsection{Application of the polynomial fit}

Overall, comparing the performance of different fitting formulae to the ejecta 
properties, we find that the \polql{} gives a comparatively better fit when all 
simulation data from all datasets are considered.
When only the \DSrefset{} and \DSheatcool{} are considered, however, 
the ejecta mass is slightly better fitted by non-polynomial fitting formula, 
Eq.~\eqref{eq:fit_Mej_Kruger}.
The implicit inclusion of mass-ratio allows the \polql{} to capture leading trends
in the behaviour of $\avd$, $\ayd$ and $\theta_{\rm RMS}$. 
For its calibration we suggest datasets with the most advanced physics i.e., 
\DSheatcool{} and \DSrefset{}. A caution must be exercised when using 
datasets computed with different physics input and at various resolutions, 
as in certain cases (e.g., $\amd$), the systematics introduced 
by these differences might obscure 
the leading trends in data.
This conclusion is supported by the analysis of the statistical behaviour 
of data from different datasets and further corroborated by the analysis 
of the individual datasets (see Appendix~\ref{app:datasets}).
In addition to the quantitative and qualitative assessments 
of the fit performance via \ssr, 
$\chid$-statistics and residual plots, 
we consider a direct application of the, 
\polql{} and compare it to the data used for calibration in Fig.~\ref{fig:ej_parspace}.
The plot shows that the behaviour of the fitting formula depends sensibly on the 
choice of datasets used for calibration, and the predictive power of the fit reduces 
when datasets with different physics (the difference in contour shapes between left and 
right column of subplots) and numerical setups are employed.
The ejecta properties, especially, mass, velocity and electron fraction depend 
strongly 
on the neutrino treatment scheme and larger number of high resolution NR simulations
with advanced treatment of neutrino emission and absorption is required to further 
constrain the statistics of ejecta properties.

\section{Remnant disk}
\label{sec:remdisk}

\begin{figure}[t]
    \centering 
    \includegraphics[width=0.48\textwidth]{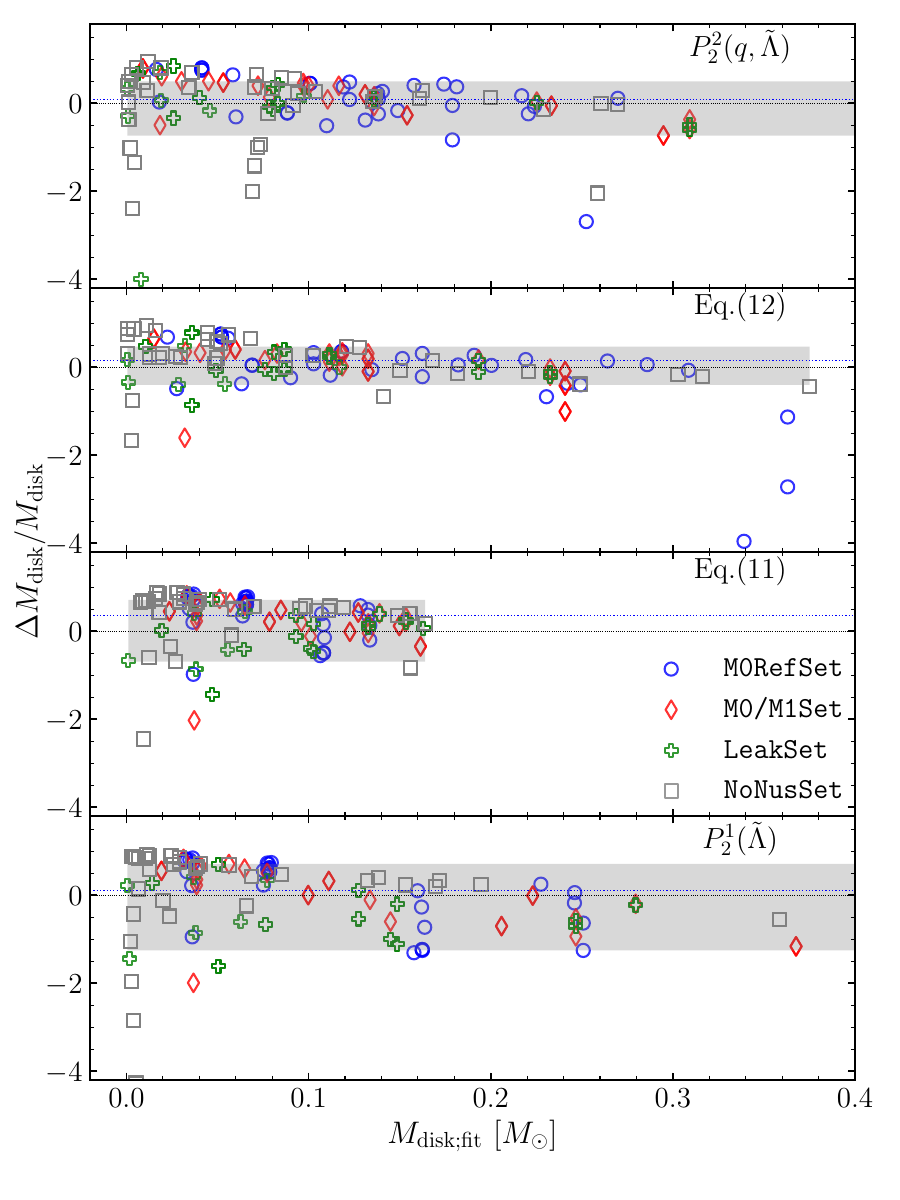}
    \caption{
        Relative differences between data and the fits of the disk mass. 
        The calibration was performed for $\log_{10}(M_{\rm disk})$ 
        using simulations from all datasets. 
        Different panels show polynomial fits in $\tilde{\Lambda}$ and $(q,\tilde{\Lambda})$, fitting formulae Eq.~\eqref{eq:fit_Mej} and Eq. \eqref{eq:fit_Mej_Kruger}. 
        The best fitting model is characterized by the lowest value of $\chi_{\nu}^2$.
        Best fitting coefficients are given in the tables in Appendix~\ref{app:coefs}.
        Here $\Delta M_{\rm disk} = M_{\rm disk} - M^{\rm fit}_{\rm disk}$.
        The fitting procedure here was based in minimizing residuals instead of $\chi_{\nu}^2$ as otherwise, the error measure adapted, Eq.~\eqref{eq:disk:mdisk_err}, would lead to the 
        fit underestimating most of datasets used.
    }
    \label{fig:stat_mdisk}
\end{figure}

\begin{figure*}[t]
    \centering 
    \includegraphics[width=0.49\textwidth]{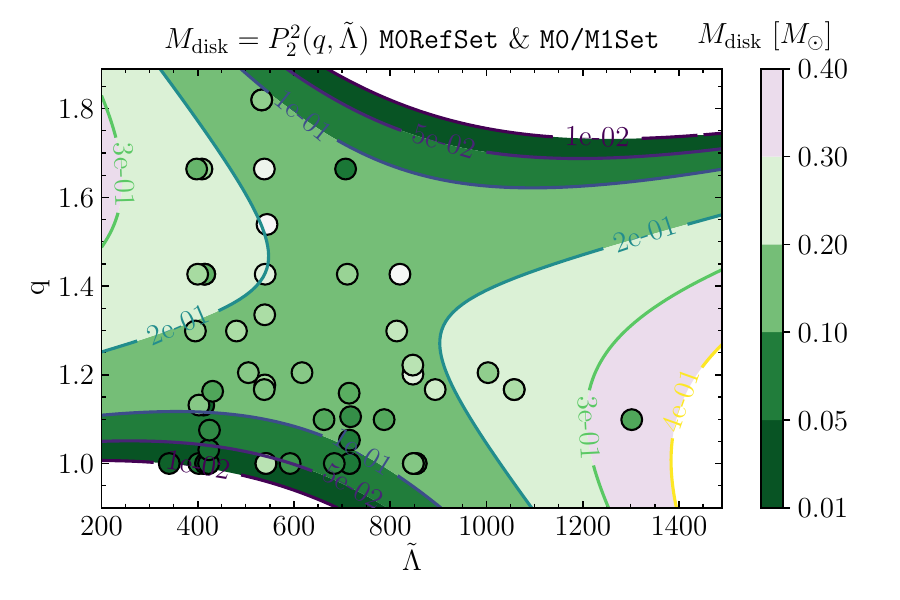}
    \includegraphics[width=0.49\textwidth]{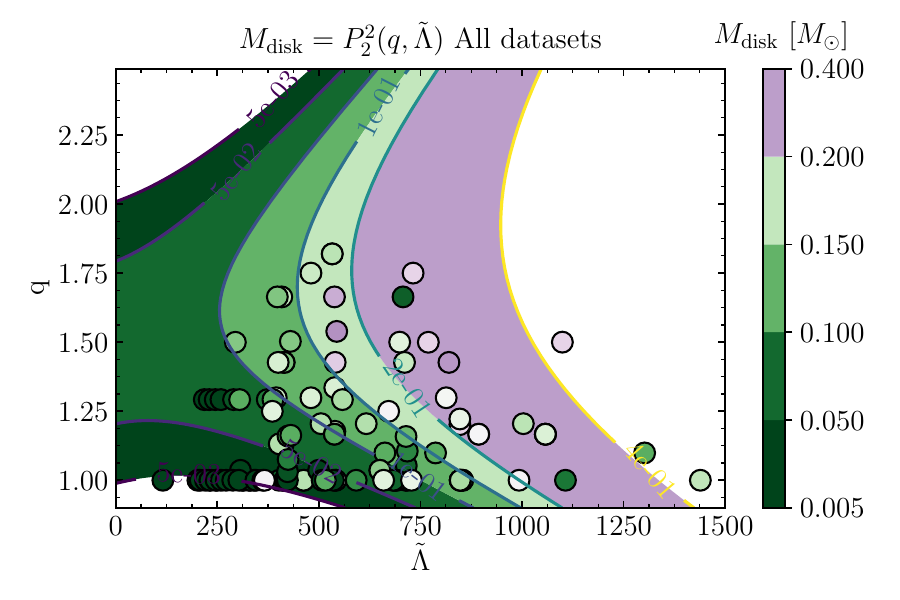}
    \caption{
        Same as Fig.~\ref{fig:ej_parspace} but for the disk mass.
        The plot shows that at low values of $q$ and $\tilde{\Lambda}$ the fit 
        is able to capture the leading trends in data. However, in the region where 
        there are fewer data preset, at high $q$ and $\tilde{\Lambda}$, the fit 
        becomes increasingly less accurate (see text).
    }
    \label{fig:mdisk_parspace}
\end{figure*}


\begin{table}[t]
    \label{tab:DiskFitchi2}
    \caption{
        Sum of squared residuals for different
        fit models for the final disk mass, $\log_{10}(M_{\rm disk})$.
    }
    \begin{tabular}{l|cccccc}
        \hline\hline
        datasets & Mean & Eq.~\eqref{eq:fit_Mdisk} & Eq.~\eqref{eq:fit_Mdisk_Kruger} & $P_2^1(\tilde{\Lambda})$ & $P_2^2(q,\tilde{\Lambda})$ \\ \hline
        \texttt{M0RefSet} & 15.11 & 13.28 & 9.96 & 13.95 & 8.81 \\
        \& \texttt{M0/M1Set} & 17.03 & 14.42 & 11.58 & 15.24 & 10.70 \\ 
        \& \texttt{LeakSet} & 54.02 & 32.56 & 17.65 & 29.72 & 19.56 \\ 
        \& \texttt{NoNusSet} & 80.47 & 45.71 & 30.06 & 44.04 & 26.73 \\ 
        \hline\hline
    \end{tabular}
    \label{tbl:fit:mdisk:chi2dofs}
\end{table}


The disk mass at the end of the simulation of models of \DSrefset{} varies from 
$0.01M_{\odot}$ to $0.3M_{\odot}$.
Within the restricted parameter space the mean value of the disk mass, 
$\overline{M_{\rm disk}}$, for models of the \DSrefset{} is 
$(0.12 \pm 0.05) M_{\odot}$ and it decreases only slightly when models 
from \DSheatcool{}, \DScool{} are added, to $(0.11\pm0.04)M_{\odot}$.
Notably, large variations in the mean value are observed when the 
parameter space is enlarged to include very asymmetric and promptly 
collapsing models. However, there is not enough models for the comparison.
While this might suggest that the disk mass depends weakly on the physical
setup of simulations, the large uncertainties in data and the fundamental
difference between the disk around a neutron star and a black hole must 
be taken into consideration.
In particular, we stress that the disk mass is estimated in different ways in the
different datasets. 
In \cite{Dietrich:2015iva,Dietrich:2016hky} the
disk is estimated only for BNS forming BH, at approximately ${\approx}
1$~ms after collapse and computing the rest mass outside the apparent 
horizon (AH). 
In \cite{Sekiguchi:2016bjd}, the disk mass is extracted at
${\approx} 30$~ms outside the AH. In \cite{Radice:2018pdn}, the disk mass is computed
as the baryonic mass outside the AH at BH formation, while for NS
remnants the criterion $\rho < 10^{13}$ g cm$^{-3}$ is used. 
In \cite{Kiuchi:2019lls} for both BH and NS outcome the $\rho < 10^{13}$ g cm$^{-3}$ 
criterion is used and time of the extraction is not specified. 
In \cite{Vincent:2019kor} the density criterion is the same, however the simulations 
are significantly shorter (${\sim 7.5}$~ms) than in other
works. Overall, we estimate that these differences can amount to a systematic factor of a few.

As fitting formulae we consider the polynomials Eqs~\eqref{eq:polyfit2} and \eqref{eq:polyfit22}, 
and the formula provided in \cite{Radice:2018pdn}:

Similarly to the mass of dynamical ejecta, the disk mass varies by 
up to an order of magnitude for, in some cases, very similar values of $q$ 
and $\tilde{\Lambda}$.
In order to simplify the fitting procedure and reproduce both, high and low 
mass tails, we consider the $\log_{10}(M_{\rm disk})$. 
Notably, the Eqs.~\eqref{eq:fit_Mdisk}-\eqref{eq:fit_Mdisk_Kruger} are segmented, 
and include constant parts.
For clarity we write the equations in the form used for fitting, that read 
\begin{equation}
\begin{split}
\log_{10}\left(\frac{M_{\rm{disk}}}{M_{\odot}}\right)_{\rm fit} & \\ =\text{max}\Big\{ -3.0,\, \log_{10}\Big(\alpha &+ \beta \tanh\Big(\frac{\tilde{\Lambda}-\gamma}{\delta}\Big)\Big)\Big\}\,,
\label{eq:fit_Mdisk}
\end{split}
\end{equation}
and the fitting formula provided in \cite{Kruger:2020gig}:
\begin{equation}
\begin{split}
\log_{10}\left(\frac{M_{\rm{disk}}}{M_{\odot}}\right)_{\rm fit} & \\
=\log_{10}(M_A) + \text{max}\Big\{ -3.30,\, &\log_{10}\Big( (\alpha C_A + \beta)^\gamma
\Big)\Big)\Big\}
\,.
\label{eq:fit_Mdisk_Kruger}
\end{split}
\end{equation}

The exact from of polynomials, \polq{} and \polql{}, used in this section than reads, 
\begin{align*}
\log_{10}\left( P_2 ^1(\tilde\Lambda) \right) &= 
\log_{10}\Big(b_0 + b_1\tilde\Lambda + b_2 \tilde\Lambda^2\Big), \\\label{eq:polyfit22log}
\log_{10}\left( P_2 ^2(q,\tilde\Lambda) \right) &= \log_{10}\Big(b_0 + b_1q + b_2\tilde\Lambda + b_3q ^2 \\
&+  b_4 q \tilde\Lambda + b_5\tilde\Lambda^2 \Big) \, .
\end{align*}
As before we opt here for the minimization of residuals in the fitting procedure.
We rank the fitting formulae performance based on the \ssr,
augmenting the dicsusion with $\chid$, computed using the error measure 
\eqref{eq:disk:mdisk_err}.

The coefficients of the polynomial regressions are reported in 
Tab.~\ref{tab:diskfit:poly}; 
the fit coefficients for 
Eq.~\eqref{eq:fit_Mdisk} and 
Eq.~\eqref{eq:fit_Mdisk_Kruger}
are reported in 
Tab.~\ref{tab:diskfit:form}.
The \ssr{} for these fits are reported in Tab.~\ref{tab:DiskFitchi2}.
As for those for the dynamical ejecta, the formulae in
Eq.~\eqref{eq:fit_Mdisk} and Eq.~\eqref{eq:fit_Mdisk_Kruger} give
ill-conditioned fits. 
Notably, we find that depending on the initial guess for coefficients 
the Eq.~\eqref{eq:fit_Mdisk_Kruger} may develop 
singularities when data from all datasets is fitted and no limitations 
are imposed upon the coefficients.
However, such non-smooth fitting functions may allow to 
capture the complex behavior in data, not reproduced by \polq{} and \polql{}

\begin{figure*}[t]
    \centering 
    \includegraphics[width=0.49\textwidth]{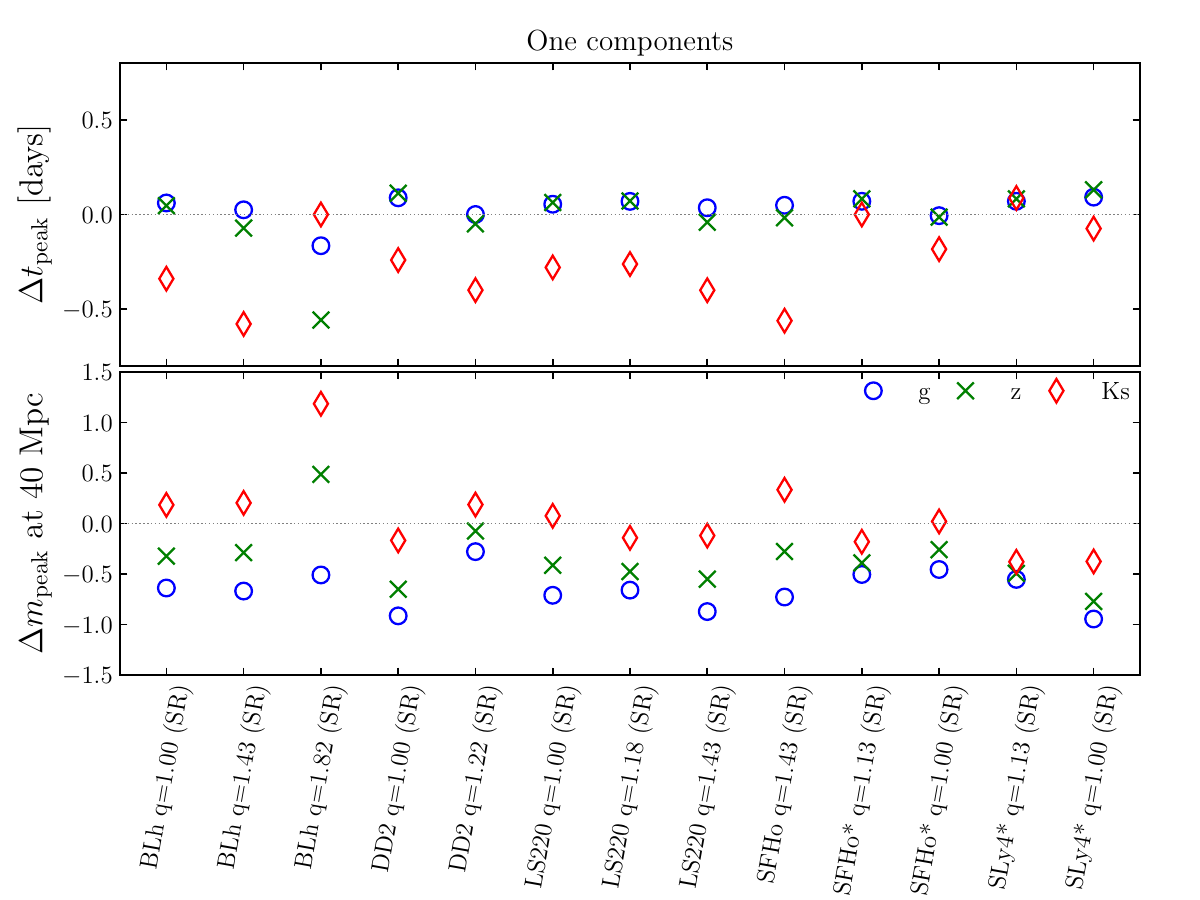}
    \includegraphics[width=0.49\textwidth]{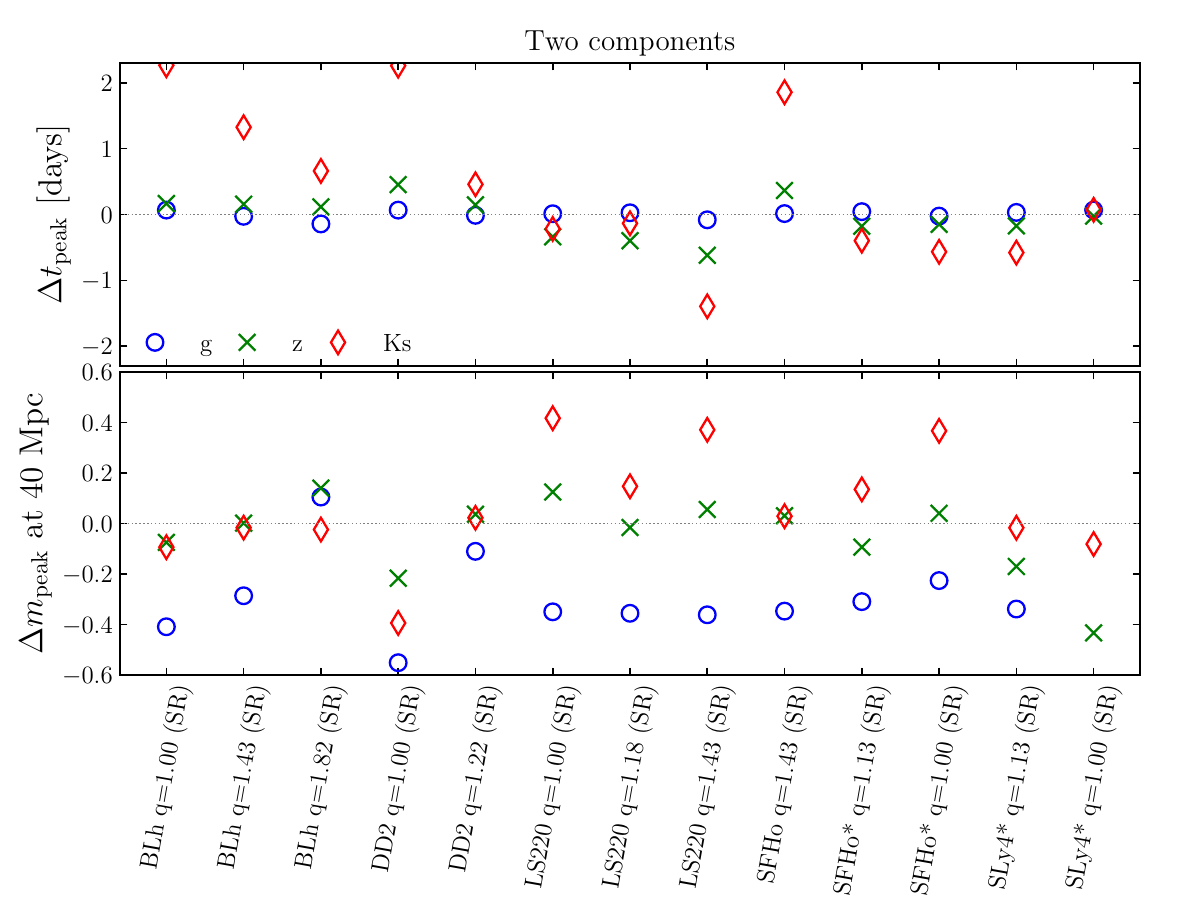}
    \caption{
        Comparison between one component light curves (\textit{left panel}) and
        two components light curves (\textit{right panel}) in $g$, $z$ and $K_s$
        bands using direct NR input or the fitting formulae for the
        dynamical ejecta and disk mass. 
        The $y-$axis displays the difference between the peak time (\emph{top panel}), 
        $\Delta t_{\rm peak} = t_{\rm peak; NR} - t_{\rm peak; fit}$, and peak magnitude, 
        $\Delta m_{\rm peak} = m_{\rm peak; NR} - m_{\rm peak; fit}$, (\emph{bottom panel});
        the $x-$axis shows selected BNS models of \DSrefset{}.
        The fits employed here are the polynomials in $(q,\tilde{\Lambda})$ used with the 
        best fitting coefficients, calibrated to \DSheatcool{} (that includes \DSrefset{}).
        The plot shows that 
        the light curves generated with the dynamical ejecta fits (one
        component) tend to underestimate the peak times and magnitudes
        of NR-informed light curves, especially in the $K_s$ band. 
        In case of dynamical ejecta and disk wind (two components) light curves, the peak
        time is less constrained ($\pm 2$~days) in the $K_s$ band, but the
        peak magnitudes is predicted more accurately $\pm0.5$~mag. }
    \label{fig:mkn_example}
\end{figure*}

Fitting the data of \DSrefset{} and combined \DSrefset{} dataset and \DSheatcool{} we 
observe that the \polql{} consistently shows the lowest \ssr{} and $\chid{}$.
Notably, the Eq.~\eqref{eq:fit_Mdisk_Kruger} gives only slightly higher values 
in both cases.
When all models from all datasets are considered, we again observe that the 
\polql{} is statistically preferred with Eq.~\eqref{eq:fit_Mdisk_Kruger} being the 
close second. 
The observed similarity in fitting formulae performance further suggests that 
indeed mass-ratio and $\tilde{\Lambda}$ allow to capture the main trends in the disk mass data.

When performing the calibration of Eq.~\eqref{eq:fit_Mdisk_Kruger} and 
Eq.~\eqref{eq:fit_Mdisk} with standard least-square method we observed that the 
result of the calibration  
depends strongly on the initial guesses for the coefficients. 
This behavior makes the use of these fitting formulae difficult from the point 
of view of the reproducible of result. 
We also note that Eqs.~\eqref{eq:fit_Mdisk}-\eqref{eq:fit_Mdisk_Kruger} include 
constant ``floor values''. The physical motivation behind these 
constants is not very clear and 
while they might help to constrain the fit behaviour at known 
limits of the parameter space, \eg, at $\Lambda\rightarrow0$, 
their applicability for all datasets may not 
be optimal.
The \polql{} fitting formula is free from aforementioned issues and allows 
for stable and reproducible fits.

In Fig.~\ref{fig:stat_mdisk} we show the relative differences
between the data and the values given by the fitting models. 
Here the relative performance of the fits can be 
inferred  
from the $67\%$ confidence level bar.
We observe that the Eq.~\eqref{eq:fit_Mdisk} cannot reproduce 
the high disk masses found in asymmetric binaries of \DSrefset{}. 
Meanwhile other fitting formulae can reproduce both the low and the 
large disk masses with varying degree of success. Notably, the fit 
with Eq~\eqref{eq:fit_Mdisk_Kruger} displays the smallest residuals, 
\ie, with the narrowest $67\%$ confidence level bar.
The second best fit here is \polql{}.
The reason why the $\chid$ for the Eq.~\eqref{eq:fit_Mdisk_Kruger} 
is larger than that for \polql{} (see Tab.~\ref{tab:DiskFitchi2}) lies in 
the error measure, Eq.~\eqref{eq:disk:mdisk_err}, that is used only 
in $\chid$ calculation. Thus, while the fit with lowest $\chid$ provides a 
better fit for lower disk masses (with tighter errors), the 
Eq.~\eqref{eq:fit_Mdisk_Kruger} gives a fit with overall smaller 
residuals.

We show the performance of the \polql{} fitting 
formula in the $q$-$\tilde{\Lambda}$ space in Fig.~\ref{fig:mdisk_parspace}.
The plot shows that certain main trends in data, e.g., higher disk mass 
in low-$q$, low-$\tilde{\Lambda}$ simulations, are reproduced by the 
fit calibrated with either combined \DSrefset{} and \DSheatcool{} or all datasets.
However, being a smooth function, the \polql{}, cannot capture the 
rapid oscillations in data.

Overall, the statistical analysis shows that the value of the disk mass 
is subjected to large uncertainties, that include systematic 
and method-of-computation uncertainties. 
The leading trends in the data appears to be captured by the 
fitting formulae that include mass-ratio and reduced tidal deformability.
This result is generally supported by the datasets separate statistical 
analysis (see Appendix~\ref{app:datasets}).
As a simple polynomial in terms of mass ration and the reduced 
tidal deformability shows similar or better residuals and $\chid$, compared 
to other fitting formulae available in the literate literature and formulated in terms of 
other binary parameters, we conclude that the former two quantities describe the 
leading trends in data.
The analysis of all datasets individually generally confirms this conclusion, 
further suggesting that both \polql{} and Eq.~\eqref{eq:fit_Mdisk_Kruger} 
perform similarly well (see Appendix~\ref{app:datasets}).

\section{Discussion}

In this paper we considered numerical relativity datasets available in the literature for
the dynamical ejecta properties and the remnant disk mass from binary neutron star
mergers. 
We performed a simple statistical analysis of the ejecta parameters that highlighted  
that the ejecta parameters are subjected to large systematic uncertainties, 
partially due to 
different treatment of neutrinos, in addition to the EOS formulations.
We also compared different fitting formulae for the ejecta properties and disk mass 
and found that fitting formulae that include the reduced tidal parameter and mass ratio can relatively well 
reproduce the leading trends in certain datasets with more uniform physics input. 
In particular, low order polynomials in these quantities provide a simple description 
of the data available and also favorably compare to the other options in terms of 
sum of squared residuals 
when only models of \DSrefset{} are considered as well or models from all datasets.
Large values of \ssr{} and $\chid{}$ as well as 
wild oscillations of fitting coefficients for a given quantity between calibrations 
(see App.~\ref{app:coefs}) further indicate the limitations on 
the ability of the set of all simulations to preserve physical information.
This calls for more 
detailed studies of error estimates in simulations containing the necessary physics.
Additionally, a larger sample of simulations with parameters more uniformly distributed 
is required as the current set available in the literature is rather limited in terms of 
mass and mass-ratio, and mostly concentrated around binaries with fiducial $1.4\Msun$ NS.
Nonetheless, since these fitting formulas are widely used for multimesseneger analyses, 
we propose the use of these polynomial models instead of other 
fitting formulae presented in the literature (and also considered 
in this work) because most of these formulae lead to ill-conditioned fits.
Specifically, we recommend the Eq.~\eqref{eq:polyfit22} calibrated with datasets with the 
most advanced physics input, \ie, \DSheatcool{} and \DSrefset{} 
(highlighted rows in Tab.~\ref{tab:dynfit:poly} and Tab.~\ref{tab:diskfit:poly})
We empathize that the application of the fitting formulae, especially polynomials, 
should be limited to the parameter space where they have been calibrated.
Additionally, while our analysis suggests that for the currently available data, the 
second order polynomials in $q$ and $\tilde{\Lambda}$ perform comparatively well, 
higher-order formulae might be necessary to capture the true physics of mergers.
We leave their exploration to future works when more simulation 
data becomes available.

When all data from all available datasets are considered, 
the fitting formulae with the best statistical performance among those considered are able to 
reproduce the dynamical ejecta velocity typically to ${\sim}50\%$, with the $68\%$ 
significance range being $\Delta \vd / \vd \in (-0.4,0.2)$. 
The electron fraction is reproduced with an accuracy of ${\sim}0.1$. 
The ejecta RMS half opening angle about the orbital plane is reproduced with an 
accuracy of ${\sim}10$~deg. 
The ejecta and disk masses, however, are rather uncertain having $(-0.8, 0.2)$ and $(-0.4, 0.2)$ 
$68\%$ significance ranges respectively. 
The smooth fitting formulae can reproduce these quantities to a factor of ${\sim}2$.

The main conclusion of this work is that the currently available data on the ejecta 
properties and disk masses from binary neutron star mergers contains large systematic 
uncertainties. 
Different treatments of EOS and neutrino transport, as well as  different resolutions, and methods of 
calculation of ejecta and disk properties lead to large systematic differences between 
various datasets.
As neutrino re-absorption is a crucial component for reliable estimates of the dynamical ejecta mass,
e.g. \cite{Wanajo:2014wha,Sekiguchi:2015dma,Perego:2017wtu,Foucart:2018gis},
it is of paramount importance to enlarge the \DSheatcool{} and refine the statistics of ejecta properties.
Additionally, 
different methodologies 
used to extract and compute these quantities contribute to the uncertainties. 
Simulations of sequences of binaries at different chirp masses could also be useful
to identify new trends in the data that cannot be currently explored.
The statistical analysis that we have performed is further subjected to 
biases as the data in different datasets span different ranges in parameter 
space of the binary. 
Considerably larger sets of simulations that cover the parameter space 
more uniformly are need to alleviate these biases.

Fitting formulae to the ejecta properties and disk mass are commonly used
to study sources of the gravitational waves and electromagnetic counterparts.
However, caution ought to be exercised when applying the fitting formulae presented
here to infer the source parameters from observations.

As an example, we discuss the impact of using our 
recommended, \polql{}, fitting formula 
for the computation of synthetic kilonova light curves as opposed to
the direct numerical relativity input\footnote{The ejecta mass, velocity and electron fraction distributions
are used to compute the light curve as in Ref.~\cite{Nedora:2019jhl}}.
We use the semi-analytic model of
Ref.~\cite{Perego:2017wtu} with one or two kilonova components,
\textit{i.e.,} the dynamical ejecta and the disk wind.
We consider a set of selected BNS models from the 
\DSrefset{} with 5 different EOS and several mass rations between $q=1.00$ and $q=1.82$.
From the \polql{} we estimate the dynamical ejecta 
mass and velocity and angle separating the low opacity polar part 
and high opacity part about the plane of the binary, using the $\theta_{\rm RMS}$
as a separation angle.
We invoke the ejecta mass-averaged RMS half opening angle to separate the low-altitude 
high opacity part and the low-opacity polar part. 
This allows us to include the change in ejecta geometry with binary parameters.
For the secular ejecta mass we assume it to be $40\%$ of
the disk mass, evaluated from the best fitting formula.
The opacities, heating rates and extrinsic parameters are kept fixed in the comparison.

The results are collected in Fig.~\ref{fig:mkn_example}, where we show
peak times and peak magnitudes for the $g$, $z$, and $K_s$ filters.
In the one component case (left panels), we find that the peak times are
reproduced on average within ${\sim}0.2$~days in the $g$
an $z$ bands, and within ${\sim}0.5$~days in $K_s$ band. The latter is
systematically underestimated. The highly asymmetric binary $q=1.8$ and BLh EOS shows
overall the largest deviations.
Peak magnitudes in the three bands computed with the fitting formulae
differ by ${\sim}0.5$~mag from the NR informed ones, reaching ${\sim}1$~mag in 
the $g$ band.
In the two component case (right panels) the peak times in the $K_s$ band
based on the best fitting formulae are more uncertain and amount to ${\sim}2$~days. 
The peak magnitude show deviations of ${\sim}\pm0.5$~mag in $z$ and $K_s$
bands. 
The reason why the peak magnitudes are more uncertain in the one component case 
lies in the complex geometry that are inherited 
in kilonova models from the numerical relativity data, 
but is not sufficiency well captured by the single parameter,
mass-averaged RMS half opening angle, considered here.
While the precise details and origin of these differences can
be related to the specific light curve model employed here, this example 
indicates the minimum systematic variation is to be expected 
in the light curve predictions when using 
our recommended fitting formula.

\begin{acknowledgments}
  We thank the anonymous referees for their comments that helped 
  us improve the manuscript. 
  We thank Erika Holmbeck for useful discussions.
  S.B., B.D. and F.Z. acknowledge support by the EU H2020 under ERC Starting
  Grant, no.~BinGraSp-714626.
  D.R.~acknowledges support from the U.S. Department of Energy, Office of Science, Division of Nuclear Physics under Award Number(s) DE-SC0021177 and from the National Science Foundation under Grant No. PHY-2011725.
  Data postprocessing was performed on the Virgo ``Tullio'' server 
  at Torino supported by INFN.
\end{acknowledgments}

\appendix

\section{Tables with fitting coefficients}
\label{app:coefs}

This appendix summarizes all fit coefficients.
Dynamical ejecta coefficients can be found in 
Tab.~\ref{tab:dynfit:poly} and 
Tab.~\ref{tab:dynfit:fit_form} for the polynomials and fitting
formulae respectively.
Disk coefficients can be found in 
Tab~\ref{tab:diskfit:poly} and 
Tab.~\ref{tab:diskfit:form} for the polynomials and fitting
formulae respectively.
The coefficients of the recommended fitting formulae, as discussed in
the conclusion, are highlighted in the tables.
Importantly, the range of the binary parameters of the 
datasets used for calibration ought to be taken into account 
when the fitting formulae are used.
The corresponding ranges are discussed in Sec.\ref{sec:method}.


\begin{table*}
    \caption{
        \label{tab:dynfit:poly}
        Dynamical ejecta properties:
        coefficients for polynomial regression of various
        quantities. Results for both
        first order and second order polynomials are reported $P_2^1(\tilde{\Lambda})$ and $P_2^2(q, \tilde{\Lambda})$
        The recommended calibration for $P_2^2(q,\Lambda)$ is highlighted.
    }
    \begin{tabular}{l|l|ccccccccc}
        \hline\hline
        Quantity &Datasets & $b_0$ & $b_1$ & $b_2$ & $b_3$ & $b_4$ & $b_5$ &  $\chid$   \\ \hline
        $\log_{10}(\md)$ & \DSrefset{} & $-3.49$ & $3.51\times10^{-3}$ & $-3.01\times10^{-6}$ & & & & 1.9  \\ 
        & \& \DSheatcool{} & $-2.40$ & $-7.11\times10^{-5}$ & $-1.60\times10^{-7}$ & & & & 18.8  \\ 
        & \& \DScool{} & $-3.37$ & $1.85\times10^{-3}$ & $-1.21\times10^{-6}$ & & & & 14.3  \\ 
        & \& \DSnone{} & $-2.53$ & $-2.03\times10^{-5}$ & $-6.74\times10^{-9}$ & & & & 46.0  \\ 
        \hline
        $\vd$ [c] &  \DSrefset{} & $4.28\times10^{-1}$ & $-8.46\times10^{-4}$ & $6.42\times10^{-7}$ & & & & 2.9  \\ 
        & \& \DSheatcool{} & $3.37\times10^{-1}$ & $-4.70\times10^{-4}$ & $3.16\times10^{-7}$ & & & & 3.2  \\ 
        & \& \DScool{} & $2.75\times10^{-1}$ & $-2.36\times10^{-4}$ & $1.39\times10^{-7}$ & & & & 6.2  \\ 
        & \& \DSnone{} & $2.50\times10^{-1}$ & $-6.66\times10^{-5}$ & $2.15\times10^{-8}$ & & & & 7.6  \\ 
        \hline
        $\yd$ & \DSrefset{} & $3.26\times10^{-1}$ & $-6.16\times10^{-4}$ & $5.70\times10^{-7}$ & & & & 42.7  \\ 
        & \& \DSheatcool{} & $1.98\times10^{-1}$ & $-3.05\times10^{-5}$ & $4.64\times10^{-8}$ & & & & 38.3  \\ 
        & \& \DScool{} & $1.45\times10^{-1}$ & $1.09\times10^{-4}$ & $-6.89\times10^{-8}$ & & & & 36.0  \\ 
        \hline
        $\athetarms$ [deg] & \DSrefset{} & $3.95\times10^{+1}$ & $-4.96\times10^{-2}$ & $5.00\times10^{-5}$ & & & & 21.2 \\ 
        & \& \DSheatcool{} & $2.41\times10^{1}$ & $7.21\times10^{-3}$ & $2.28\times10^{-6}$ & & & & 18.3  \\ 
        & \& \DScool{} & $1.44\times10^{1}$ & $3.42\times10^{-2}$ & $-1.81\times10^{-5}$ & & & & 14.1  \\ 
        \hline\hline
        $\log_{10}(\md)$ & \DSrefset{} & $0.436$ & $-2.75$ & $-6.18\times10^{-3}$ & $2.75\times10^{-1}$ & $4.78\times10^{-3}$ & $3.96\times10^{-7}$ & 1.2  \\ 
\rowcolor{lightgray} & \& \DSheatcool{} & $-1.32$ & $-3.82\times10^{-1}$ & $-4.47\times10^{-3}$ & $-3.39\times10^{-1}$ & $3.21\times10^{-3}$ & $4.31\times10^{-7}$ & 20.8  \\ 
        & \& \DScool{} & $-6.96$ & $5.26$ & $7.84\times10^{-4}$ & $-1.71$ & $5.69\times10^{-4}$ & $-9.09\times10^{-7}$ & 7.9  \\ 
        & \& \DSnone{} & $-6.01$ & $4.91$ & $-1.24\times10^{-3}$ & $-1.57$ & $1.00\times10^{-3}$ & $2.77\times10^{-8}$ & 17.9  \\
        \hline
        $\vd$ [c] & \DSrefset{} & $6.10\times10^{-1}$ & $-1.12\times10^{-1}$ & $-1.04\times10^{-3}$ & $-6.56\times10^{-2}$ & $3.56\times10^{-4}$ & $4.25\times10^{-7}$ & 0.9  \\ 
\rowcolor{lightgray} & \& \DSheatcool{} & $5.94\times10^{-1}$ & $-1.48\times10^{-1}$ & $-8.62\times10^{-4}$ & $-5.02\times10^{-2}$ & $3.25\times10^{-4}$ & $3.16\times10^{-7}$ & 1.6  \\ 
        & \& \DScool{} & $2.55\times10^{-1}$ & $1.88\times10^{-1}$ & $-4.44\times10^{-4}$ & $-1.46\times10^{-1}$ & $1.87\times10^{-4}$ & $1.38\times10^{-7}$ & 5.3  \\ 
        & \& \DSnone{} & $3.46\times10^{-1}$ & $-8.11\times10^{-2}$ & $-8.11\times10^{-5}$ & $-3.67\times10^{-3}$ & $8.89\times10^{-6}$ & $1.99\times10^{-8}$ & 7.0  \\ 
        \hline
        $\yd$ & \DSrefset{} & $-3.49\times10^{-2}$ & $3.01\times10^{-1}$ & $5.55\times10^{-4}$ & $-1.52\times10^{-1}$ & $-2.06\times10^{-4}$ & $-2.44\times10^{-7}$ & 8.7  \\ 
\rowcolor{lightgray} & \& \DSheatcool{} & $2.55\times10^{-1}$ & $3.83\times10^{-2}$ & $2.36\times10^{-4}$ & $-6.66\times10^{-2}$ & $-1.92\times10^{-4}$ & $-1.86\times10^{-8}$ & 9.6  \\ 
        & \& \DScool{} & $-2.58\times10^{-1}$ & $6.33\times10^{-1}$ & $5.02\times10^{-4}$ & $-2.41\times10^{-1}$ & $-3.04\times10^{-4}$ & $-1.25\times10^{-7}$ & 24.8 \\
        \hline
        $\athetarms$ [deg] & \DSrefset{} & $-7.79\times10^{1}$ & $1.38\times10^{2}$ & $1.30\times10^{-1}$ & $-5.50\times10^{1}$ & $-3.33\times10^{-2}$ & $-7.25\times10^{-5}$ & 4.4  \\ 
\rowcolor{lightgray} & \& \DSheatcool{} & $-5.61\times10^{1}$ & $1.29\times10^{2}$ & $6.88\times10^{-2}$ & $-5.27\times10^{1}$ & $-2.72\times10^{-2}$ & $-2.78\times10^{-5}$ & 4.1  \\ 
        & \& \DScool{} & $-1.06\times10^{2}$ & $1.79\times10^{2}$ & $1.11\times10^{-1}$ & $-6.10\times10^{+1}$ & $-6.59\times10^{-2}$ & $-2.48\times10^{-5}$ & 8.5  \\
        \hline\hline
    \end{tabular}
\end{table*}

\begin{table*}
    \caption{
        \label{tab:dynfit:fit_form}
        Dynamical ejecta properties:
        coefficients for the fitting formulae discussed in the text for various datasets.}
    \begin{tabular}{l|l|l|ccccccc}
        \hline\hline
        Quantity&Fit & Datasets & $\alpha$ & $\beta$ & $\gamma$ & $\delta$ & $n$ &  $\chid$  \\
        \hline
        $\log_{10}(\md)$ & Eq.~\eqref{eq:fit_Mej}  & \DSrefset{} & $9.662\times10^{-2}$ & $1.037$ & $5.034$ & $-8.316$ & $2.432\times10^{-1}$ & 1.6  \\ 
        &      & \& \DSheatcool{} & $-1.004\times10^{-1}$ & $-4.403\times10^{-1}$ & $-6.452\times10^{-1}$ & $2.696\times10^{-1}$ & $3.222\times10^{-1}$ & 6.0  \\ 
        &      & \& \DScool{} & $-1.067\times10^{-1}$ & $-1.651$ & $2.806$ & $2.784$ & $3.013\times10^{-1}$ & 13.6  \\ 
        &      & \& \DSnone{} & $9.429\times10^{-2}$ & $-7.036\times10^{-1}$ & $2.121$ & $-1.026$ & $5.328\times10^{-1}$ & 29.9  \\ 
        \hline
        $\log_{10}(\md)$ & Eq.~\eqref{eq:fit_Mej_Kruger}  & \DSrefset{} & $-2.361\times10^{-3}$ & $2.750\times10^{-2}$ & $-8.573\times10^{-2}$ & & $1.279$ & 1.4  \\ 
        &      & \& \DSheatcool{} & $-1.261\times10^{-3}$ & $1.449\times10^{-2}$ & $-4.715\times10^{-2}$ & & $1.306$ & 5.1  \\ 
        &      & \& \DScool{} & $-1.153\times10^{-3}$ & $1.285\times10^{-2}$ & $-4.164\times10^{-2}$ & & $1.339$ & 6.1 \\ 
        &      & \& \DSnone{} & $-3.351\times10^{-4}$ & $2.697\times10^{-3}$ & $-9.738\times10^{-3}$ & & $1.729$ & 20.0  \\ 
        \hline
        $\vd$ [c] & Eq.~\eqref{eq:fit_vej}& \DSrefset{} & $-7.242\times10^{-1}$ & $1.279$ & $-1.537$ & & & 1.2  \\ 
        &      & \& \DSheatcool{} & $-5.631\times10^{-01}$ & $1.109$ & $-1.186$ & & & 2.3  \\ 
        &      & \& \DScool{} & $-4.007\times10^{-1}$ & $9.164\times10^{-1}$ & $-6.881\times10^{-1}$ & & & 6.0  \\ 
        &      & \& \DSnone{} & $-3.627\times10^{-1}$ & $8.191\times10^{-1}$ & $-1.128$ & & & 6.8  \\ 
        \hline\hline
    \end{tabular}
\end{table*}


\begin{table*}
    \caption{
        \label{tab:diskfit:poly}
        Disk mass:
        coefficients for polynomial regression of various
        quantities. Results for both
        first order and second order polynomials are reported $P_2^1(\tilde{\Lambda})$ and $P_2^2(q, \tilde{\Lambda})$
        The recommended calibration for $P_2^2(q,\Lambda)$ is highlighted.
        Note, that here the $\log_{10}$ of the RHS of respective polynomials is considered.
    }
    \begin{tabular}{l|ccccccccc}
        \hline\hline
        Datasets & $b_0$ & $b_1$ & $b_2$ & $b_3$ & $b_4$ & $b_5$ &  $\chid$   \\
        \hline
        \texttt{M0RefSet} & $-3.62\times10^{-1}$ & $1.42\times10^{-3}$ & $-9.60\times10^{-7}$ & & & & 477.8  \\ 
        \& \texttt{M0/M1Set} & $-1.76\times10^{-1}$ & $7.50\times10^{-4}$ & $-4.01\times10^{-7}$ & & & & 323.6  \\ 
        \& \texttt{LeakSet} & $3.53\times10^{-2}$ & $-3.12\times10^{-4}$ & $6.88\times10^{-7}$ & & & & 37.3  \\ 
        \& \texttt{NoNusSet} & $1.05\times10^{-2}$ & $-1.44\times10^{-4}$ & $4.99\times10^{-7}$ & & & & 61.0  \\ 
        \hline
        \texttt{M0RefSet} & $-1.80$ & $2.44$ & $7.87\times10^{-4}$ & $-6.78\times10^{-1}$ & $-8.08\times10^{-4}$ & $2.80\times10^{-7}$ & 8.8  \\ 
        \rowcolor{lightgray} \& \texttt{M0/M1Set} & $-1.85$ & $2.59$ & $7.07\times10^{-4}$ & $-7.33\times10^{-1}$ & $-8.08\times10^{-4}$ & $2.75\times10^{-7}$ & 26.6  \\ 
        \& \texttt{LeakSet} & $-1.26$ & $1.76$ & $3.51\times10^{-4}$ & $-4.82\times10^{-1}$ & $-5.20\times10^{-4}$ & $3.68\times10^{-7}$ & 18.9  \\ 
        \& \texttt{NoNusSet} & $-5.10\times10^{-1}$ & $7.78\times10^{-1}$ & $-3.29\times10^{-4}$ & $-2.60\times10^{-1}$ & $2.33\times10^{-4}$ & $2.92\times10^{-7}$ & 18.1 \\ 
        \hline\hline
    \end{tabular}
\end{table*}

\begin{table*}
    \caption{
        \label{tab:diskfit:form}
        Disk mass: coefficients for the fitting formulae discussed in the text for various datasets. }
    \begin{tabular}{l|l|ccccccc}
        \hline\hline
        Fit & Datasets & $\alpha$ & $\beta$ & $\gamma$ & $\delta$  & $\chid$  \\
        \hline
        Eq.~\eqref{eq:fit_Mdisk} & \DSrefset{} & $9.958\times10^{-2}$ & $5.346\times10^{-2}$ & $4.793\times10^{2}$ & $6.106$ & 298.4  \\ 
        & \& \DSheatcool{} & $1.026\times10^{-1}$ & $5.095\times10^{-2}$ & $4.710\times10^{2}$ & $5.351\times10^{-1}$ & 203.0  \\ 
        & \& \DScool{} & $7.677\times10^{-2}$ & $8.752\times10^{-2}$ & $5.835\times10^{2}$ & $3.429\times10^{2}$ & 105.1  \\
        & \& \DSnone{} & $7.656\times10^{-2}$ & $8.765\times10^{-2}$ & $5.840\times10^{2}$ & $3.474\times10^{2}$ & 75.0  \\ 
        \hline
        Eq.~\eqref{eq:fit_Mdisk_Kruger} & \DSrefset{} & $-6.852$ & $1.191$ & $1.346$ & & 25.5  \\ 
        & \& \DSheatcool{} & $-7.184$ & $1.303$ & $1.613$ & & 55.4  \\ 
        & \& \DScool{} & $-5.217$ & $0.902$ & $1.090$ & & 18.8  \\ 
        & \& \DSnone{} & $-8.963$ & $1.769$ & $2.841$ & & 39.3  \\
        \hline
        \hline\hline
    \end{tabular}
\end{table*}

\section{Statistics for individual datasets}
\label{app:datasets}

\begin{figure*}[t]
    \centering 
    \includegraphics[width=0.49\textwidth]{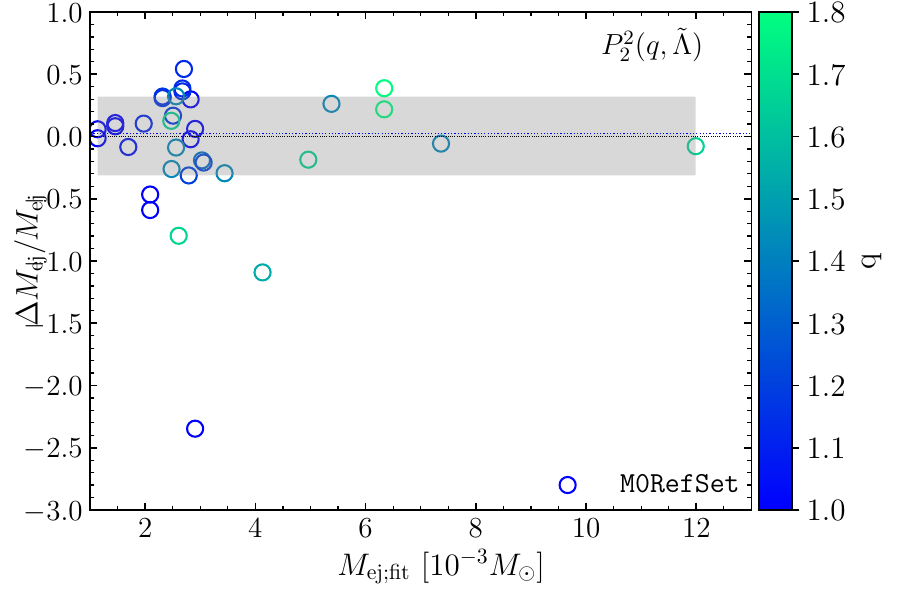}
    \includegraphics[width=0.49\textwidth]{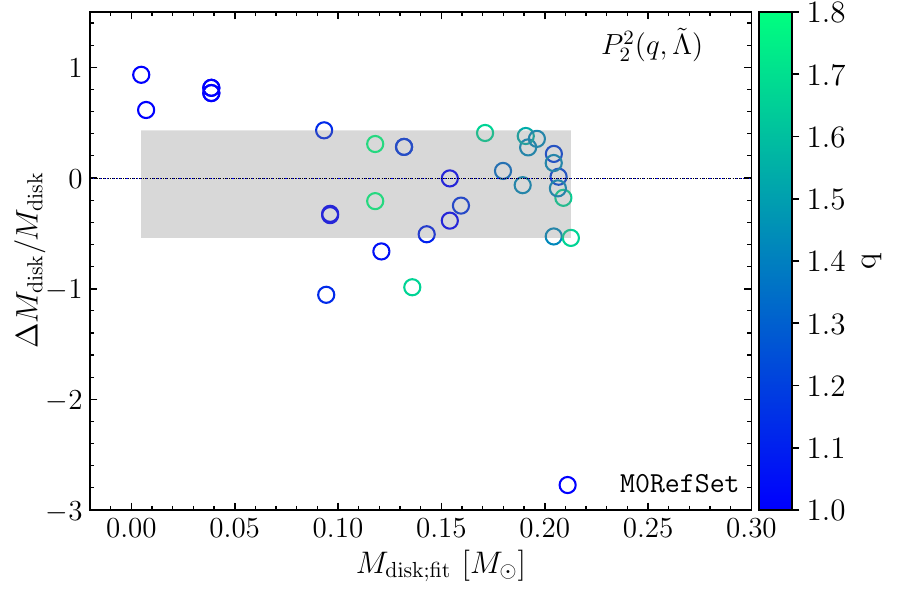}
    \includegraphics[width=0.49\textwidth]{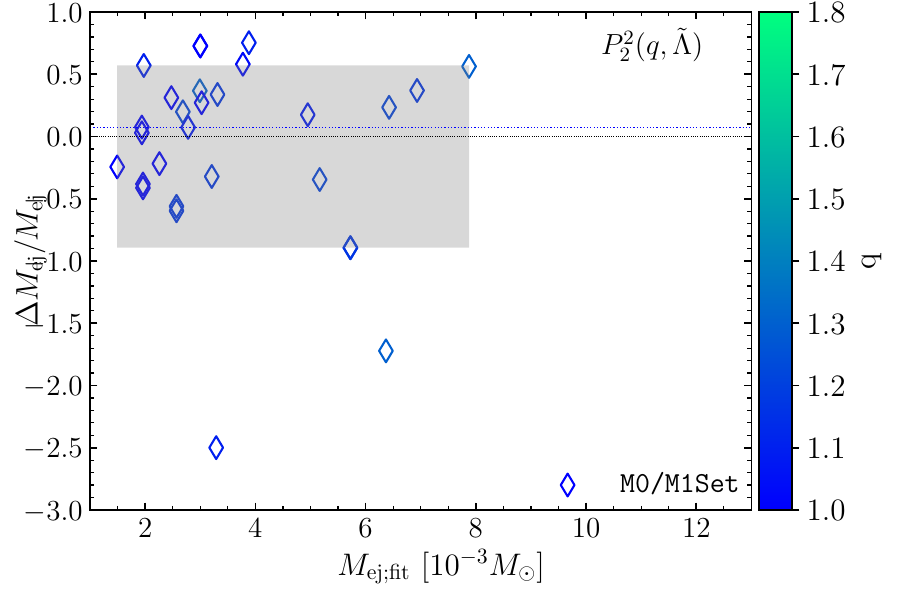}
    \includegraphics[width=0.49\textwidth]{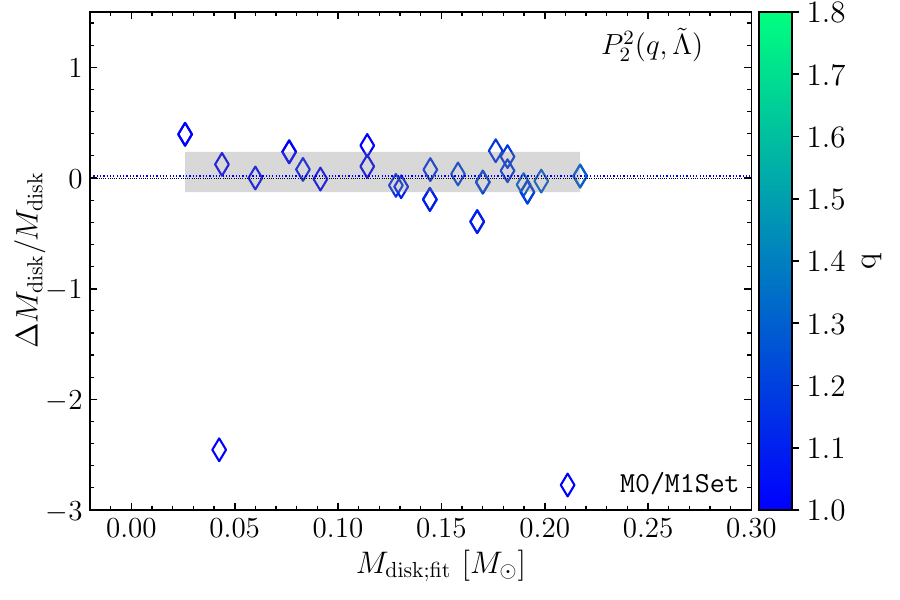}
    \includegraphics[width=0.49\textwidth]{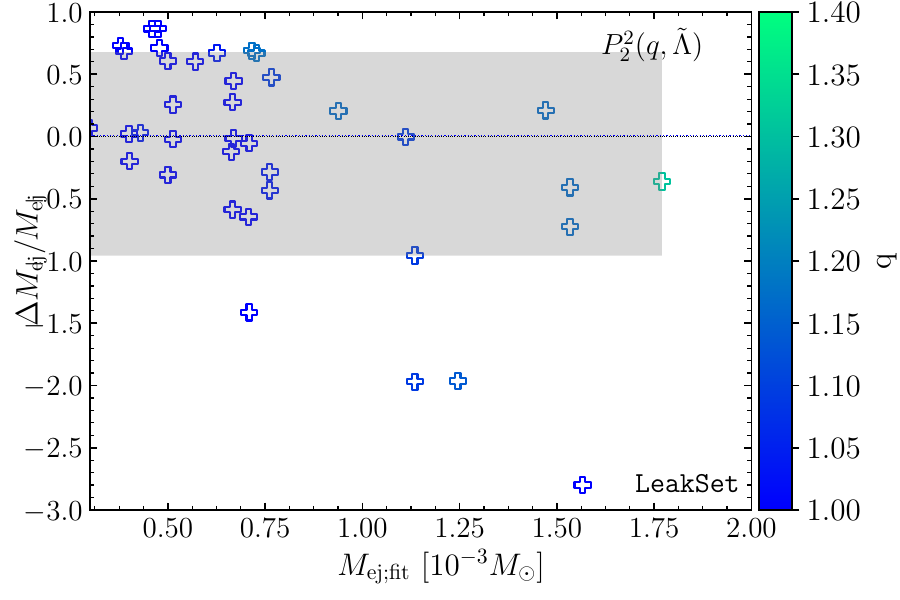}
    \includegraphics[width=0.49\textwidth]{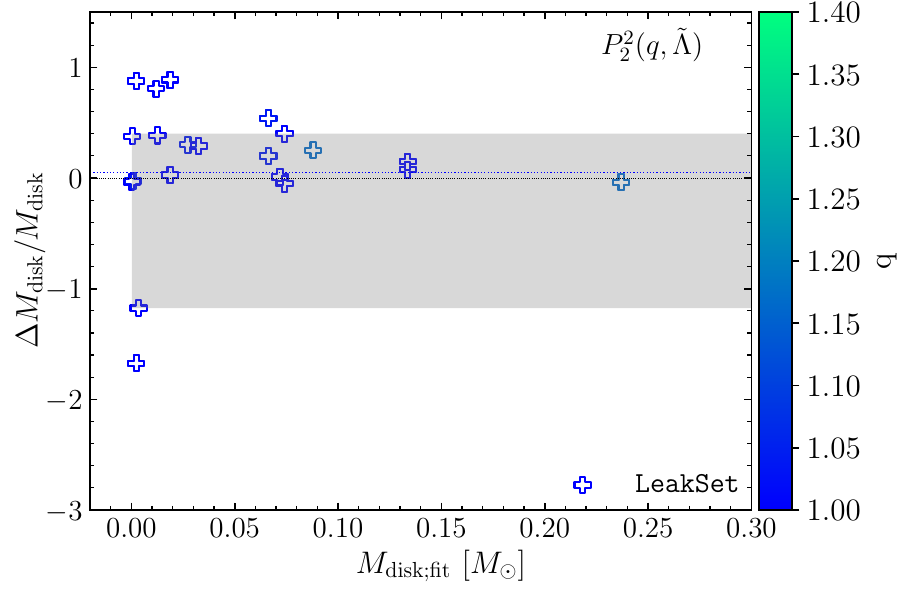}
    \includegraphics[width=0.49\textwidth]{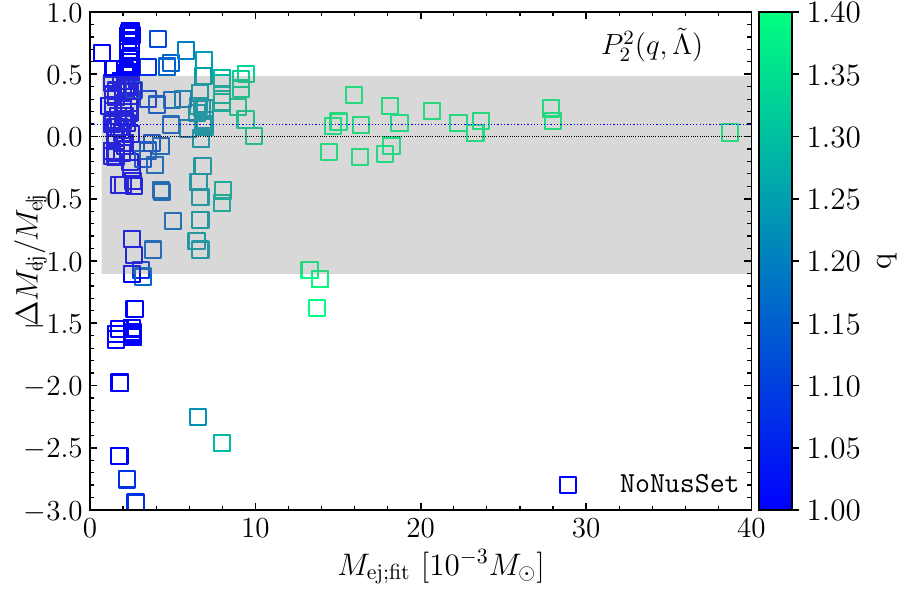}
    \includegraphics[width=0.49\textwidth]{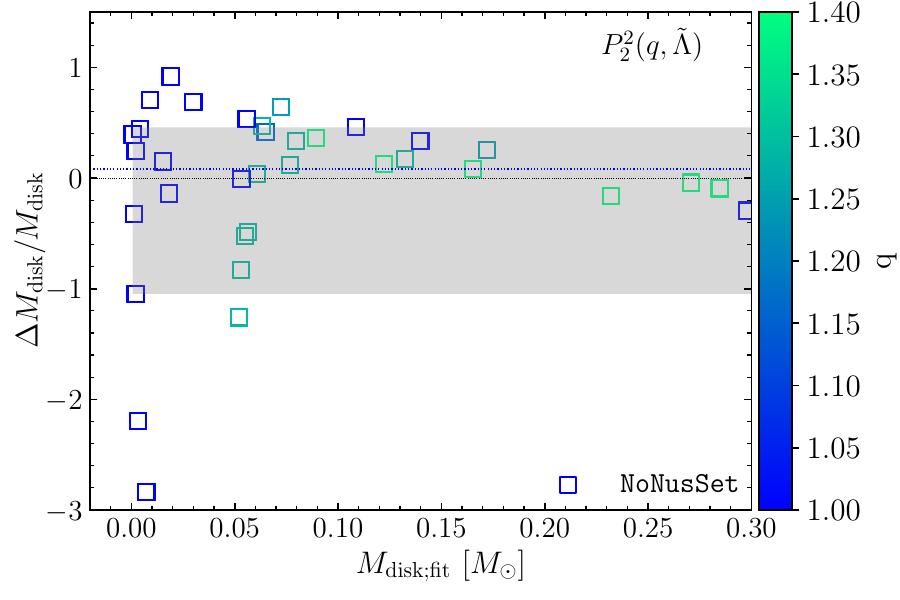}
    \caption{
        Comparison between the data and values obtained from the fitting formula \polql{} 
        for the ejecta mass (\emph{left column of plots}) and disk mass (\emph{right column of plots}).
        The plot is similar to the \ref{fig:ejecta:dyn:m} and \ref{fig:stat_mdisk}, but 
        instead of showing the result for the combined dataset with all models, in each 
        panel only one dataset is used to calibrate the \polql{} fitting formula. 
    }
    \label{fig:mej_mdisk_ind_dsets}
\end{figure*}

In this appendix we discuss the \ssr{} and $\chid$ statistics of all 
fitting formulae dataset-vise instead of adding them up, as was done in the main text.
In Tab.~\ref{tbl:fit:ejecta:chi2dofsallInd} we compare the different fits 
for the dynamical ejecta properties and disk mass in terms of the 
\ssr, 
and in the Fig.~\ref{fig:mej_mdisk_ind_dsets} we show the residuals of 
the \polql{}, with different calibrations for ejecta 
mass and disk mass.

Regarding the ejecta mass we find that \polql{} and Eq.~\eqref{eq:fit_Mej_Kruger} 
display the lowest \ssr{}. While for \DSrefset{} and \DSnone{} the \polql{} is 
preferred, for the other two datasets, the Eq.~\eqref{eq:fit_Mej_Kruger} 
gives slightly lower \ssr{}. Additionally we note that  
the datasets that are 
more uniform in their physics and method, \eg, \DSrefset{} and 
\DScool{} display the lowest $\chid$.
The largest $\chid$ is found for the \DSheatcool{}, that 
incorporates both, models with M1 and leakage+M0 neutrino schemes. 
Notably, \eqref{eq:fit_Mej} shows similar values of $\chid$ for 
\DSheatcool{}, \DSrefset{} and \DScool{}.
Fig.~\ref{fig:mej_mdisk_ind_dsets} also shows that the \polql{} reproduces 
the models of \DSheatcool{}, \DScool{} and \DSnone{} less robustly than those of \DSrefset{}.
In part this is due to the limited $\tilde{\Lambda}$ range of models of \DSrefset{} 
and fixed chirp mass, and in part it hints at the systematic uncertainties due to different 
phsysics setup of simulations.

For the ejecta velocity, the \polql{} gives the lowest \ssr{} for 
all datasets. Meanwhile, the largest $\chid$ is found 
for the \DScool{} across all fitting formulae. 
This might be attributed to the systematic uncertainties that pure 
leakage neutrino scheme introduces for models with different outcomes, \eg,
prompt collapse and stable remnants. 

With respect to ejecta electron fraction and RMS half opening angle, 
\polql{} gives significantly lower \ssr{} than \polq{} and 
the difference in $\chid$ are small. Notably, for the 
$\ayd$, the $\chid$ is similar between the \DSrefset{} and \DSheatcool{}. 
This indicates the consistency in neutrino reabsorption effects on the ejecta 
composition in these datasets.

For the disk mass we find 
that the lowest \ssr{} is given \polql{} for all datasets. 
The largest $\chid$ is found for \DSrefset{} and the smallest 
for \DSheatcool{}.
The reason for that is largely due to the error measure that we use to compute the $\chid$. 
For instance, if we employ the error bars for the \DSrefset{} individually for each model 
(See Tab.~1 in \cite{Nedora:2020pak}), we obtain $\chid\sim1$. However, this information 
is not available for other datasets and the uniform error measure, 
Eq.~\eqref{eq:disk:mdisk_err} was chosen for consistency. 
The Fig.~\ref{fig:mej_mdisk_ind_dsets} shows that indeed, the \polql{} 
reproduces the data of \DSheatcool{} much better than of any other dataset, 
with lower residuals. This can be attributed to the fact that models 
of \DSheatcool{} span a more narrow range in mass ratios and does not include prompt collapse 
models that can lead to either massive disks in very asymmetric binaries \cite{Bernuzzi:2020txg}
or a negligible disks in equal mass but massive ones \cite{Radice:2018pdn}.

Overall, the datasets-wise statistical analysis of ejecta properties and disk mass 
shows the same qualitative picture reported in the main text.


\begin{table}[t]
    \caption{
        Sum of squared residuals, \ssr, for different fitting models for the dynamical 
        ejecta properties and disk mass (see descriptions of the 
        Tab.~\ref{tbl:fit:ejecta:chi2dofsall} and Tab.~\ref{tbl:fit:mdisk:chi2dofs}). 
        Here datasets are not added, but considered individually. 
    }
    \label{tbl:fit:ejecta:chi2dofsallInd}
    \scalebox{0.88}{
        \begin{tabular}{l|l|cccccc}
            \hline\hline
            $\log_{10}(\md)$ & Datasets & $ N $ & Mean & Eq.~\eqref{eq:fit_Mej} & Eq.~\eqref{eq:fit_Mej_Kruger} & $P_2^1(\tilde{\Lambda})$ & $P_2^2(q,\tilde{\Lambda})$ \\ \hline
            & \DSrefset{} & 34 & 2.57 & 1.65 & 1.40 & 2.43 & 0.97 \\ 
            & \DSheatcool{}  & 30 & 5.56 & 3.32 & 4.35 & 5.04 & 4.49 \\ 
            & \DScool{}  & 42 & 12.70 & 10.24 & 9.73 & 11.36 & 10.64 \\ 
            & \DSnone{}  & 165 & 43.74 & 25.78 & 25.57 & 43.35 & 20.40 \\ 
            \hline\hline
            $\langle v_{\rm ej}\rangle$ & Datasets & $ N $ & Mean & Eq.~\eqref{eq:fit_vej} & & $P_2^1(\tilde{\Lambda})$ & $P_2^2(q,\tilde{\Lambda})$ \\ \hline
            & \DSrefset{} & 34 & 0.04 & 0.02 & & 0.04 & 0.01 \\ 
            & \DSheatcool{}  & 27 & 0.04 & 0.03 & & 0.03 & 0.02 \\ 
            & \DScool{}  & 42 & 0.17 & 0.17 & & 0.16 & 0.14 \\ 
            & \DSnone{}  & 143 & 0.40 & 0.30 & & 0.33 & 0.29 \\ 
            \hline\hline
            $\langle Y_{\rm e}\rangle$ & Datasets & $ N $ & Mean &  & & $P_2^1(\tilde{\Lambda})$ & $P_2^2(q,\tilde{\Lambda})$ \\ \hline
            & \DSrefset{} & 34 & 0.14 & &  & 0.13 & 0.02 \\ 
            & \DSheatcool{} & 30 & 0.05 & &  & 0.05 & 0.02 \\ 
            & \DScool{} & 35 & 0.04 & & & 0.04 & 0.04 \\ 
            \hline\hline
            $\langle \theta_{\rm RMS}\rangle$ & Datasets & $ N $ & Mean & & & $P_2^1(\tilde{\Lambda})$ & $P_2^2(q,\tilde{\Lambda})$ \\ \hline
            & \DSrefset{} & 34 & 2775 & & & 2631 & 498 \\ 
            & \DSheatcool{}  & 7 & 54 & & & 49 & 30 \\ 
            & \DScool{}  & 35 & 1355 & & & 1048 & 843 \\ 
            \hline\hline
            $M_{\text{disk}}$ & Dataset & $N$ & Mean & Eq.~\eqref{eq:fit_Mdisk} & Eq.~\eqref{eq:fit_Mdisk_Kruger} & $P_2^1(\tilde{\Lambda})$ & $P_2^2(q,\tilde{\Lambda})$ \\ \hline
            & \DSrefset{} & 31 & 15.11 & 13.28 & 9.96 & 13.95 & 8.81 \\ 
            & \DSheatcool{} & 23 & 1.88 & 0.93 & 0.59 & 1.18 & 0.44 \\ 
            & \DScool{} & 26 & 28.33 & 14.42 & 6.85 & 6.73 & 6.36 \\ 
            & \DSnone{} & 39 & 25.66 & 12.08 & 4.24 & 10.37 & 5.14 \\ 
            \hline\hline
        \end{tabular}
    }
\end{table}

\section{Effect of the error measure on the fitting procedure results}
\label{app:fiterror}

In the main text, 
the comparison between different fitting 
formulae and their respective calibration is performed using residuals (\ssr{}). 
Additionally we discuss the $\chid$ using the error measures 
found in the literature.  

In this appendix we investigate how different error measures and different 
criteria for fitting procedure affect the result. We focus 
on the models of \DSrefset{} only, for which we have errors estimated 
directly from the numerical relativity (NR) simulations performed at 
different resolutions (See Table 1 in \cite{Nedora:2020pak}). 
We also limit the analysis to the \polql{} fitting formula.
We consider three approaches: (i) minimizing the residuals, 
(ii) miminizing the $\chid$ with the default errors, discussed in the main
text and (iii) minimizing $\chid$ with the NR-informed errors.
For the (i) we compute two $\chid$, computed for both error measures.

\begin{figure*}[t]
    \centering 
    \includegraphics[width=0.49\textwidth]{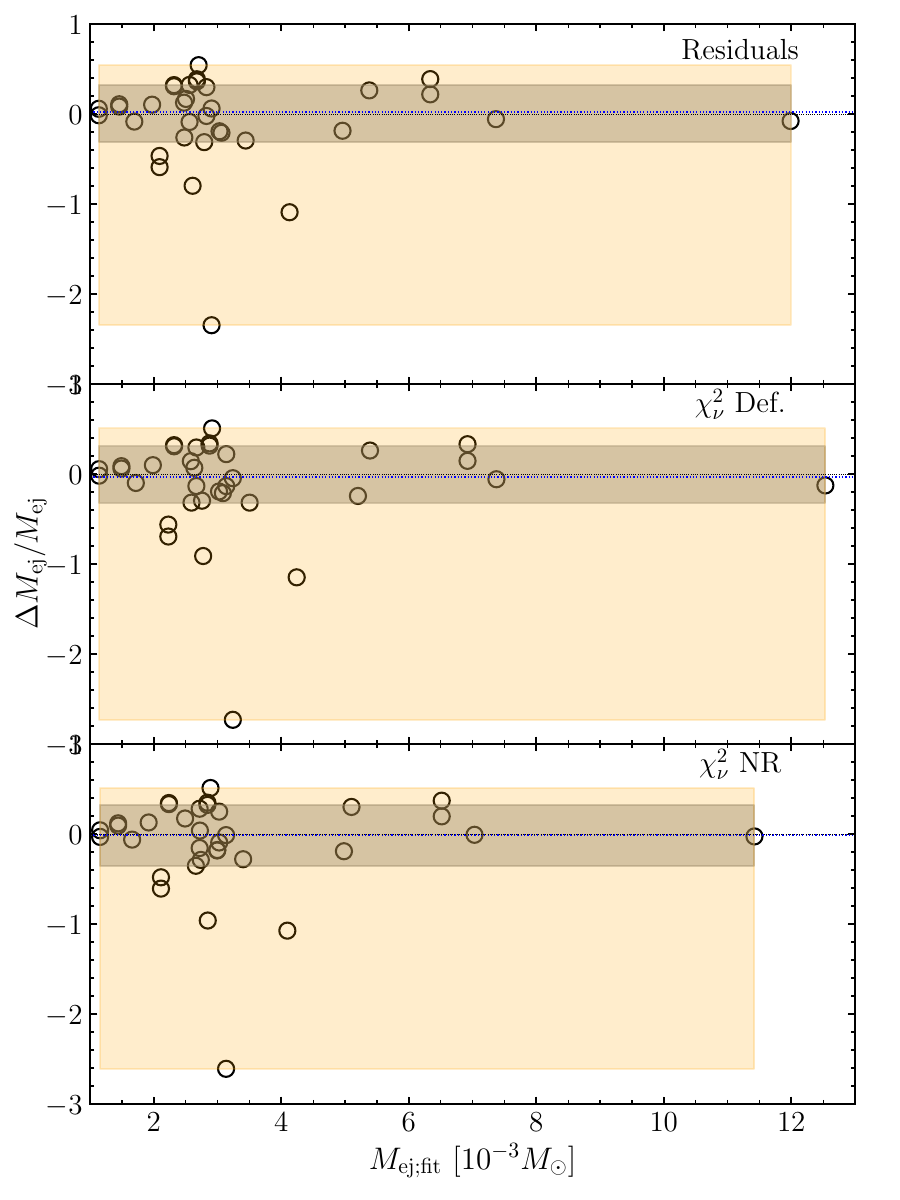}
    \includegraphics[width=0.49\textwidth]{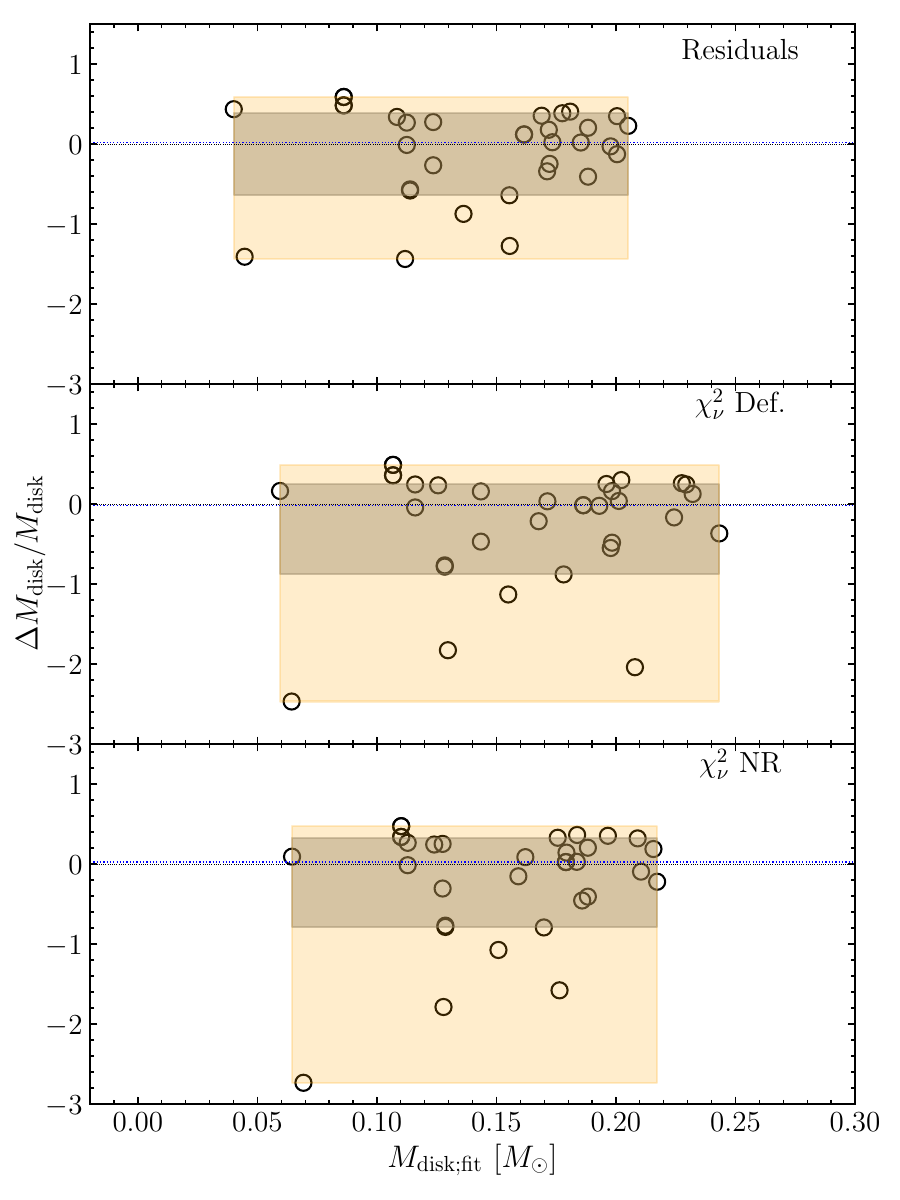}
    \caption{
        Effects of different calibration methods for \polql{} fit for models of \DSrefset{} for 
        the ejecta mass (\emph{left panel}) and disk mass (\emph{right panel}).
        We consider the minimization of residuals, $\chid$ with default error measures 
        (Eqs.~\eqref{eq:ejecta:mej_err},\eqref{eq:disk:mdisk_err}), and the errors estimated
        from multi-resolution analysis in \cite{Nedora:2020pak}.
        Gray / pastel shaded regions mark the $1\sigma\,(68\%)$ / $2\sigma\,(95.4\%)$ confidence
        intervals.
    }
    \label{fig:mej_mdisk_errors}
\end{figure*}

For the $\amd$ we observe that for (i) the $\chid$ increases by almost $3$ 
orders of magnitude when employing the NR-informed errors from 
$1.17$ to $563.92$. Meanwhile, the difference in the quality of the fit computed 
with minimization of $\chid$ using these two error measures, \ie, (ii) and (iii),
changes only slightly, as Fig.~\ref{fig:mej_mdisk_errors} shows. 
As expected, the extrema of $\Delta M_{\rm ej} / M_{\rm ej}$ are the lowest, 
$(-2.34, 0.54)$
when the residuals are minimize. However, even when $\chid$ is minimized, the 
increase in extrema is not significant (with respect to the overall fit performance):
$(-2.73, 0.50)$ for default error measure and 
$(-2.60, 0.51)$ for NR-informed errors.

For the $\avd$, we observe no difference between the fit calibrated minimizing 
residuals (i) or minimizing $\chid$ with default error (ii), as the error measure 
is a constant value. However, the increase in $\chid$ amounts to an order of 
a magnitude from $0.9$ to $9.3$. When the NR-informed error is used the fit changes 
slightly at the lower tail of the velocity with the decrease in $\chid$ to $3.3$.

Similar behaviour is observed for the $\ayd$ and $\athetarms$ as the error measure 
for these quantities are also constants.

For the $M_{\rm disk}$ we observe the similar picture as for the $\amd$. 
For (i) the $\chid$ increases by ${\gtrsim}3$ orders of magnitude from 
$1.5$ to $192$. However, the difference in the fit quantitative performance 
with minimization of $\chid$ using the two error measures, \ie, (ii) and (iii),
remains within data points' error bars (see Fig.~\ref{fig:mej_mdisk_errors}, 
left panel).

\begin{figure*}[t]
    \centering 
    \includegraphics[width=0.49\textwidth]{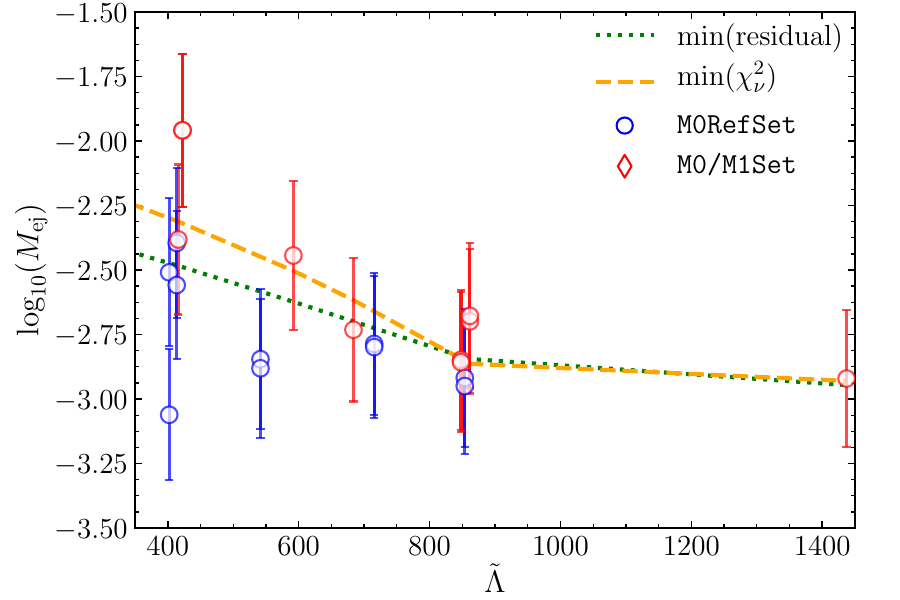}
    \includegraphics[width=0.49\textwidth]{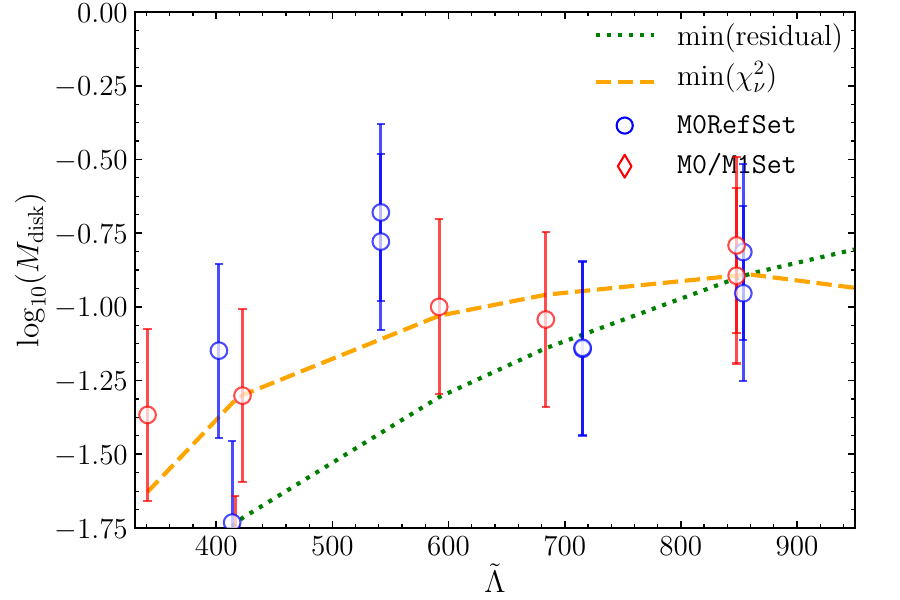}
    \caption{
        Visual representation of the \polql{} fit for ejecta mass (\emph{left panel}) 
        and disk mass (\emph{right panel}). The fits are calibrated with 
        \DSheatcool{} and \DSrefset{}, however, only models with $q=1$ are plotted. 
        The fit calibration is done either minimizing residuals or $\chid$. In the latter 
        case, the default errors are used (and also plotted) namely, 
        Eq.~\eqref{eq:ejecta:mej_err} and Eq.~\eqref{eq:disk:mdisk_err}
    }
    \label{fig:mej_mdisk_errors_q1}
\end{figure*}

For a fixed $q=1$ the performance of the \polql{} is shown in 
Fig.~\ref{fig:mej_mdisk_errors_q1}. 
We observe that the largest difference in both cases amounts to $0.25$ 
in $\log_{10}$ of the respective quantity. The fit computed minimizing the 
$\chid$ gives higher values across the 
considered range of $\tilde{\Lambda}$. 

The qualitative behavior of the fits remain, however, unchanged.

That the outcome of the fit calibration depends on the choice of 
the error measure only when this measure is biased. Otherwise, it is equivalent 
to minimizing residuals, as is the case for all quantities considered except masses. 
Regarding the latter, while the qualitative behavior of the fit appears to be 
independent of the minimization technique, the quantitative difference 
is present. 
The error measures considered in the main text are motivated by the 
finite-resolution errors found in numerical simulations \cite{Radice:2018pdn}.
However, their use for the statistical analysis of different datasets 
performed with different physics and numerical setups might not be optimal. 
This was our motivation to minimize residuals in the fitting formulae analysis 
in the main text. 
Employing a more physically and statistically motivated error measure in 
future analysis, when larger sample of data is available, would lead to better 
constrained fits. 

\bibliography{refs20210912}

\end{document}